\documentclass[aps,nobibnotes,twocolumn,superscriptaddress,showkeys,thelongbibliography]{revtex4-2}

\usepackage{dcolumn}

\usepackage{etoolbox}

\usepackage{graphicx}
\usepackage{enumerate}
\usepackage{amsmath}
\usepackage{amssymb}  
\usepackage{mathtools}
\usepackage{bm}  
\usepackage{pdfsync}  
\usepackage{hyperref}
\usepackage{physics}
\usepackage{natbib}
\usepackage{comment}
\usepackage{float}
\usepackage{xcolor}

\usepackage{amsthm}

\begin{document}

\title{Quantum random walks on d-regular graphs with Haar-random coin operators}
\author{Alice C. Quillen}
\email{aquillen@ur.rochester.edu}
\affiliation{Department of Physics and Astronomy, University of Rochester,  Rochester, NY, 14627, USA}
\affiliation{Visitor at the Max Planck Institute for the Physics of Complex Systems}
\date{\today}

\begin{abstract}
With unitary coin operator that is a random matrix drawn from a uniform distribution with respect to the Haar measure, we construct a variant of a discrete quantum random walk using a d-regular undirected connected simple graph.  With each step of the walk, the coin operator random matrix is drawn independently from the distribution.  The expectation value over the distribution of random unitaries for the associated quantum channel gives a depolarization channel for the coin subspace and resembles the associated classical random walk where the direction a walker steps depends upon the outcome of a fair coin or balanced d-sided dice. Remarkably, despite the fact that the averaged channel depolarizes the coin subspace, measurements in the vertex subspace can be designed that would reveal information about the initial quantum state, even after many iterations of the channel. We illustrate with examples of quantum walks on Cayley graphs of Abelian groups, such as the cycle and hypercube graphs.  For Cayley graphs of Abelian groups, the averaged channel is dephasing in the Fourier basis. These quantum walks are examples of bipartite strongly interacting systems, where one subsystem is strongly perturbed, yet information about the initial state in the other subsystem is potentially measurable forever.   Due to its decoherence, a quantum random walk with a Haar-random coin would not be useful in search algorithms but could aid in understanding quantum systems with strongly perturbed subsystems. 

\end{abstract}

\maketitle
\tableofcontents

\section{Introduction}

For quantum states that are linked via an undirected graph, 
 a {\it quantum random walk} is often described as a product of two non-commuting local unitary operators that is then iterated, in analogy with a classical random walk \citep{Aharonov_1993,Watrous_2001,Kempe_2003}. 
One of the operators plays the role of a coin  and the other conditionally causes transitions between  quantum states, playing the role of a walker that chooses its direction based on the outcome of a tossed coin or the throw of a d-sided dice. 
The basis states of the quantum system describe the walker position and the state of the coin. 
Quantum walks can aid in search algorithms \citep{Childs_2004,Portugal_2018} and 
multi-particle quantum walks are capable of universal quantum computation \citep{Childs_2013}.  
For a review see \citet{VenegasAndraca_2012}. 

An {\it open quantum random walk} is similar to a quantum random walk as it also involves local quantum operations but instead of unitary operations applied to a state-vector, more general types of iterated operations are applied to a density operator  \citep{Attal_2015,Carbone_2016}.  
With a coin subspace that decoheres, an open random walk on the line can have position variance $\sigma^2 \propto t$  that grows linearly with time or walk iteration number $t$, like that of a classical random walk \citep{Brun_2003} and unlike that of a coherent quantum walk where $\sigma^2 \propto t^2$ \citep{Aharonov_1993}.

In quantum mechanics, values obtained through measurement have probability dependent on the measurement operator and state vector. 
A quantum random walk or open quantum random walk is deterministic as the iteratively applied unitary operator or Lindbladian superoperator and resulting quantum state are not functions of a probability distribution.  
In contrast, a {\it quantum random circuit} consists of a product of local unitary operators, each operating on a couple of qubits or qudits, with some of them chosen
identically and independently from a distribution \citep{Emerson_2003,Fisher_2023}. 
If quantum random circuits are iterated, they can be described as {\it ergodic quantum processes} \citep{Movassagh_2021}.  Ergodic quantum processes are a class of stochastic quantum operations that includes random and repeated quantum interactions \citep{Bougron_2022} and collision models \citep{Chisholm_2021,Ciccarello_2022,Stanzione_2025} which can describe interactions between a quantum 
system and one or more thermal reservoirs.
If noise or coupling to a thermal bath is modeled as 
 a specified completely positive trace preserving 
map then it can be called a stochastic quantum channel even though the system is described 
 with a fixed superoperator and is deterministic \citep{Graydon_2022,Stefanak_2026}.

 Quantum random walks that include variations in the choices of operators,  or operators 
 that are chosen from a distribution (e.g.,  \citealt{Whitfield_2010}) can be described as 
 having a {\it Markov coin}  \citep{Ahlbrecht_2011} or a Markov shift operator.  A 
 Markov coin usually causes decoherence   \citep{Ahlbrecht_2011} but certain types of variations 
 in the operators can allow a quantum walk to 
 illustrate the phenomenon known as Anderson localization \citep{Ghosh_2014, Derevyanko_2018}.

 In this manuscript we focus on discrete time quantum random walks where 
  the applied coin unitary operators are independently drawn from a distribution  
  during each iteration of the walk.  
We modify the Hadamard-coin quantum random walk on the line  
 \citep{Aharonov_1993,Aharonov_2001} 
 and higher dimensional ones \citep{Watrous_2001,Mackay_2002} 
 so that the coin operator is chosen identically and independently from a distribution. 
At each iteration of the walk, a different operator is chosen from this distribution. 
To facilitate analytical computation we choose a distribution that is uniform with respect to
the Haar measure.  This allows us to compute a mean or averaged quantum channel which is a quantum channel computed from the expectation value of the map over the distribution of coin operators. 
A quantum random walk with a Haar-random coin operator can be considered a limiting and special 
case of a quantum system interacting with a high temperature reservoir. 

\subsection{Some notation} 

We work with finite dimensional Hilbert spaces. 
We use notation ${\cal L} ( {\mathbb C}^{d}) $ to refer to the set of linear operators that act
on the $d$-dimensional complex vector space ${\mathbb C}^d$.  
The associated group of unitary operators is $U(d)$.   Many of our quantum walks use 
a bipartite quantum space  ${\mathbb C}^d \otimes {\mathbb C}^N$ that consists of a tensor product of $d$-dimensional space, representing a coin, and an $N$-dimensional space, representing vertices of a graph.  
We refer to the $d$-dimensional subspace with subscript $A$ and that in the $N$-dimensional
subspace with subscript $B$. 
We use $\hat I_d$ to represent the identity operator on the $d$-dimensional subspace 
and $\hat I_N$ to represent the identity operator on the $N$-dimensional subspace.  

The density operator $\rho \in {\cal L} ({\mathbb C}^{dN}) $ in the  space 
of linear operators operating on the bipartite space ${\mathbb C}^{dN}$ 
is a positive semi-definite Hermitian operator with a trace of 1.   
By {\it quantum channel} ${\cal E}(\rho)$ we mean a trace preserving completely positive linear map operating on the space of density operators.     
The quantum channel itself is also a linear operator (often called a superoperator). 
Since we work with a bipartite tensor product space, we can take the partial trace of a density operator.  
When tracing over the $d$-dimensional subspace and giving a density 
operator in the $N$-dimensional subspace, we write $\rho_B = \tr_A \rho$.   
When tracing over the $N$-dimensional  
 subspace and giving a density operator in the $d$-dimensional subspace, we write $\rho_A = \tr_B \rho$. 

In the nomenclature of quantum channels, a channel  is  {\it ergodic} if it 
contains a unique eigenvector  $\rho_*$ where  ${\cal E}(\rho_*) = \rho_*$  (a fixed point) 
with eigenvalue 1 and with $\rho_*$ in the space of density operators (positive operators with trace of 1) 
 \citep{Burgarth_2013}.
 
When labelling quantum states in an $N$-dimensional space,  it is sometimes convenient to index starting from 0, with indices ranging from 0 to $N-1$ and other times it is convenient to index from 1, 
ranging from 1 to $N$.   In this manuscript, we use both conventions. 
Vertices and edges of a graph are labelled with natural numbers (and in some cases including zero). 

The symbol $\mu_H$ represents a distribution of unitary operators in $U(N)$ that is uniform with respect to the Haar measure.  We refer to an operator $\hat V \in \mu_H$ as {\it Haar-random}. 

\section{Quantum random walk on a d-regular graph}
\label{sec:dreg}

Following \citet{Aharonov_2001}, we generate a quantum random walk from a $d$-regular 
undirected simple and connected graph $G(V, E)$ with a set $V$ of $N$ vertices  and a set $E$ of edges. 
Here $d$-regular means that 
each vertex or node has $d$ edges, and simple means the graph lacks loops connecting a vertex to itself 
and lacks multiple edges connecting the same pair of vertices.   The quantum walk is implemented on 
 a bipartite quantum space ${\mathbb C}^d \otimes {\mathbb C}^N$. 
The quantum walk is described by iterating a unitary operator 
\begin{align}
\hat U  = \Lambda_S (\hat C \otimes \hat I_N) \label{eqn:hatU}
\end{align}
where $\Lambda_S \in {\cal L}({\mathbb C}^{dN})$ is a controlled shift operator and $\hat C \in {\cal L}({\mathbb C}^d) $ is a coin operator that operates
on the $d$-dimensional subspace. 

In this section we index starting from 1 in both $d$ and $N$-dimensional subspaces. 
For each vertex $v\in V$ we assume that the $d$-regular undirected graph 
allows construction of a set of $d$ different index functions 
$w(j,v)$ \citep{Aharonov_2001,Kempe_2003} for $j \in \{1, 2, \ldots, d \}$ 
where $j$ uniquely labels the 
edges that contain the vertex $v$.  
Specifically the function $w(j,v)$ gives the vertex at the other end 
of the $j$-th edge that is connected to the vertex $v$.  
For each $j$ the function $w(j,v)$ should give a permutation that permutes all  vertices.   
So as to generate a unitary 
transformation using these functions, 
we require that  the set $\{ w(j,v) :  v \in V\} = V$  for all $j$ and so that 
 $w(j,v) = w(j, v')$ iff $v = v'$. 
For each $j$ we have a unique vertex 
 $w(j,v) \in V$ such that $(v, w(j,v)) \in E$ is an edge of the graph and this is true for every vertex $v$. 
All edges in $E$ should be present in the set $ \{ (v, w(j,v)) : v \in V, j \in \{1, 2, \ldots, d \} \}$.  
  
To represent traversal along the edge connecting vertex $v$ to vertex $w(j,v)$, 
it is convenient to define the operator 
\begin{align}
\hat W_{j} = \sum_{v=1}^N \ket{w(j,v)} \bra{v}, \label{eqn:Wop}
\end{align}
where the sum is over all vertices and $j \in \{1, 2, \ldots, d\}$. The conditions placed 
upon the graph and index functions $w(j,v)$ ensure that $\hat W_j$ is unitary. 

\begin{figure}[htbp]
\includegraphics[width=3truein]{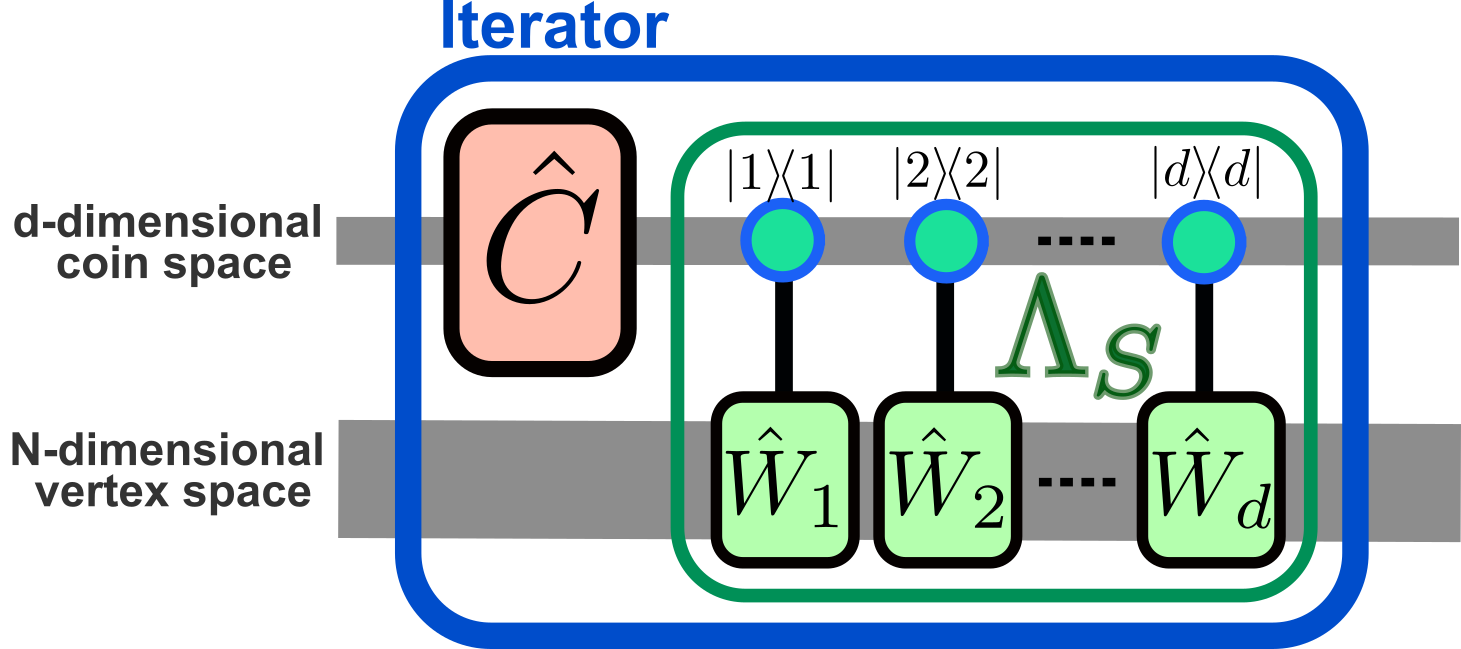}
\caption{A quantum circuit illustration of a discrete quantum random walk based on a $d$-regular graph 
that takes place on a quantum space that is  
the tensor product of a $d$-dimensional coin space and an $N$ dimensional vertex space. 
The controlled shift operator $\Lambda_S $ (outlined in green and defined in equation \ref{eqn:LambdaS0}) 
permutes the vertices with operators $\hat W_j$ (defined in equation \ref{eqn:Wop}) depending upon 
the graph edge that is selected via one of the states in the coined subspace. 
The coin operator $\hat C$ operates on the coin-subspace where it can induce superposition.  
The pair of operations is outlined in blue and is iterated to carry out the walk. 
\label{fig:illust}}
\end{figure}

With the $w(j,v)$ index functions and associated operators $\hat W_j$, we define the controlled shift unitary operator 
\begin{align}
\Lambda_S  &\equiv \sum_{j=1}^d \ket{j}\bra{j} \otimes  \sum_{v = 1}^{N} \ket{w(j,v)}\bra{v}  \nonumber \\
& =  \sum_{j=1}^d \ket{j}\bra{j} \otimes \hat W_j 
 \label{eqn:LambdaS0}.
\end{align}
Each $\ket{j}\bra{j}$ is a projection operator onto the $j$-th edge in the coin subspace. 
The controlled shift operator permutes the vertices depending upon 
the edge that is selected with the projection operator $\ket{j}\bra{j}$. 
Figure \ref{fig:illust} illustrates the quantum walk as a quantum circuit.  
The controlled shift operator can also be written as a product of controlled operators 
as shown in the illustration;  
 $\Lambda_s = \prod_{j=1}^d ( \ket{j}\bra{j} \otimes W_j + (\hat I_d - \ket{j}\bra{j} ) \otimes \hat I_N)$. 

Not all $d$-regular graphs allow construction of functions $w(j,v)$ that are permutations. 
An undirected Cayley graph of a finite Abelian group allows such a construction using generators 
of the group to give the edges of the graph \citep{Aharonov_2001}. 
A larger coin space is needed for  
 construction of a quantum walk for any undirected $d$-regular graph \citep{Watrous_2001}.

\subsubsection{Example 1: The Hadamard coin quantum walk on the cycle graph} 
\label{sec:Hcycle}

We give an example similar to the quantum walk on the line \citep{Aharonov_1993}. 
Instead of an infinite dimensional line, the graph is the cycle graph with $N$ vertices (e.g., following \cite{Aharonov_2001}).
Each vertex has two edges so the graph degree is $d=2$.   The quantum space is $2 \times N$ dimensional. 
We label the vertices so that $v$ is an integer in $\{ 0, 1, \ldots , N-1 \}$ so that we can use the modulo 
function in the conditional shift operator. 
The coin space consists of two states,  $\ket{0}$ and $ \ket{1}$. 
The index function
\begin{align}
w(j,v)_\text{cycle} = \begin{cases} 
v+1 \text{ mod } N & \text{ if } j=0 \\
v-1 \text{ mod } N & \text{ if } j=1
\end{cases} .
\end{align}
The operator associated with the edges of the graph 
\begin{align}
\hat W_{j,\text{cycle}} = \sum_{v=0}^{N-1} \ket{v +1 \text{ mod } N } \bra{v}
\end{align}
is equivalent to a Heisenberg shift operator. 
The quantum walk on the cycle graph is given by the product 
\begin{align}
\hat U_\text{cycle} =  \Lambda_{S,\text{cycle}}   (\hat C_\text{cycle}\otimes \hat I_N ) 
\end{align}
with shift operator 
\begin{align}
\Lambda_{S,\text{cycle}} &= \ket{0}\bra{0} \otimes \sum_{v=0}^{N-1} \ket{(v+1) \text{ mod } N} \bra{j} \nonumber \\
& \ \ \ +  \ket{1}\bra{1} \otimes \sum_{v=0}^{N-1} \ket{(v-1) \text{ mod } N} \bra{j}. 
\label{eqn:LS_cycle} 
\end{align}
For this quantum walk, a popular coin operator is a Hadamard operator \citep{Aharonov_1993}
\begin{align}
\hat C_\text{cycle}  = \frac{1}{\sqrt{2}} \begin{pmatrix} 1 & 1 \\ 1 & -1 \\\end{pmatrix} =\frac{1}{\sqrt{2}}  \sum_{j,k=0}^1 (-1)^{jk} \ket{j}\bra{k} . 
\end{align}

\subsubsection{Example 2: The Grover coin quantum walk on the hypercube graph}

We give an example for a quantum walk on the hypercube \citep{Kempe_2003,Shenvi_2003}.
For the $d$-dimensional hypercube, with $N = 2^d$ vertices, the $N$-dimensional 
space describing vertices is equivalent to the tensor product of $d$ two state quantum systems or $d$ qubits.  
A vertex on the hypercube can be labelled with its binary string $v = v_1 v_2 v_3 \dots v_d$ 
where each digit in the string $v_j \in \{0, 1 \}$ with index $j$ ranging from 1 to $d$. 
Equivalently $v$ is an integer ranging from $0$ to $2^d -1$. 
Instead of a 2 state coin, we have $d$ possible coin states, so the coin is more like a $d$-sided dice. 
The full quantum space for the quantum walk is ${\mathbb C}^d \otimes {\mathbb C}^N$. 
For the coin state we let $j$ range from 1 to $d$.   
The function $w(j,v)$ consists of applying a logical NOT operation applied to the $j$-th digit of the 
binary string labelling a vertex $v$. 
In other words 
\begin{align}
w(j,v_1 v_2 \ldots v_d)_\text{hypercube} = v_1 v_2  \ldots \text{NOT} (v_j) \ldots v_d .
\end{align} 
This implies that $\ket{w(j,v)} \bra{v} = \hat X_j \ket{v}\bra{v} $ where 
 $\hat X_j$ is the Pauli X operator operating on the $j$-th qubit. 
 
The quantum walk on the hypercube is given by the product 
\begin{align}
\hat U_\text{hypercube} =  \Lambda_{S,\text{hypercube}}  (\hat C_\text{hypercube}\otimes \hat I_N )
\end{align}
with shift operator 
\begin{align}
\Lambda_{S,\text{hypercube}} = \sum_{j=1}^d\ket{j}\bra{j} \otimes \hat X_j .
\label{eqn:LS_hypercube}
\end{align}
For the quantum walk on the hypercube, 
the coin operator $\hat C_\text{hypercube}$ can be a discrete Fourier transform operator \citep{Watrous_2001}
or a Grover operator;  
\begin{align}
\hat C_\text{hypercube}  
 = 2\ket{y} \bra{y} - \hat I_d 
 \end{align}
  \citep{Kempe_2003,Shenvi_2003}
with superposition state 
\begin{align}
\ket{y} \equiv \frac{1}{\sqrt{d}} \sum_{j=1}^d \ket{j}.\end{align} 

\section{A quantum walk with a Markov coin that is Haar-random}
\label{sec:Haar}

We modify the quantum random walk by choosing a coin operator $\hat V$
 from the distribution $\mu_H \in U(d)$ 
that is uniform with respect to the Haar measure.  
The coin operator 
\begin{align}
\hat C_{\tilde V} \equiv \hat V \otimes \hat I_N 
\text{ with } \hat V \in \mu_H  \label{eqn:CV}.
\end{align} 
In each application of the $\hat C_{\tilde V}$ we choose a different and independent random unitary $\hat V$ in $U(d)$. 
We use a tilde superscript on $V$ to denote an operator that is based on a distribution for $\hat V$. 
We refer to the coin operator as {\it Haar-random} to denote that it is drawn 
from a distribution that is uniform with respect to the Haar measure.  

The shift operator (equation \ref{eqn:LambdaS0}) can be implemented as a unitary quantum channel 
operating on a density operator $\rho \in {\cal L} ({\mathbb C}^{dN}) $ via 
\begin{align}
{\cal E}_S(\rho) \equiv  \Lambda_S \rho \Lambda_{S}^\dagger. \label{eqn:ES_def}
\end{align}
The coin random channel based on $\hat C_{\tilde V}$ in equation \ref{eqn:CV}, and
with $\hat V$ drawn from a distribution,  is a unitary channel 
\begin{align}
{\cal E}_{\tilde V}(\rho) \equiv ( \hat V \otimes \hat I_N)\,  \rho\, ( \hat V^\dagger \otimes \hat I_N)
\text{ with } \hat V \in \mu_H .  \label{eqn:EV_def}
\end{align}
The random matrix coined quantum walk consists of iterating the consecutive application  of the two channels 
\begin{align}
{\cal E}_{S\tilde V} (\rho) &\equiv {\cal E}_S(\cal E_{\tilde V} (\rho)) \nonumber \\
&= 
 \Lambda_S ( \hat V \otimes \hat I_N)\,  \rho\, ( \hat V^\dagger \otimes \hat I_N) \Lambda_S^\dagger 
 \label{eqn:ESV_def}
\end{align}
with $\hat V \in \mu_H$. 

\subsection{The averaged quantum channel}
\label{sec:ave}

To compute an expectation value integrating over possible values for coin operator $\hat V \in \mu_H$, 
we use an application of Corollary 13 by \citet{Mele_2024} applied to a two state system: 
For any operator $O \in {\cal L} ({\mathbb C}^d )$ (the space of linear operators that operate on a $d$-state system)
the expectation value 
\begin{align}
{\mathbb E}_{U \in \mu_H}   \left[ U O U^\dagger \right] = \tr O \frac{\hat I_d}{d}  .  \label{eqn:Mele50b}
\end{align}
(via equation 50 by \citet{Mele_2024}) where $\mu_H$ is the distribution of unitaries in $U(d)$ that is uniform 
according to the Haar measure.   

We compute the expectation value of the Haar-random coin quantum channel (defined in equation \ref{eqn:EV_def})  by 
averaging over all possible unitary operators drawn 
from the Haar uniform distribution 
\begin{align}
\bar {\cal E}_V (\rho) & \equiv {\mathbb  E}_{\hat V \in \mu_H} {\cal E}_{\tilde V} (\rho) \nonumber \\
 & = {\mathbb  E}_{\hat V \in \mu_H}  \left[ ( \hat V \otimes \hat I_N)\,  \rho\, ( \hat V^\dagger \otimes \hat I_N)   \right]  \nonumber \\ 
& =  \frac{\hat I_d}{d} \otimes \tr_A \rho. 
 \label{eqn:Adepol}
\end{align}
We have applied equation \ref{eqn:Mele50b} to the $d$-dimensional subspace.  
We use an overline to denote an averaged channel. 
The averaged channel {\it depolarizes} within the coin space. 

Taking the expectation value over the Haar-random component (applying  
equation \ref{eqn:Mele50b} to the Haar-random coin quantum walk (the channel of equation \ref{eqn:ESV_def}), 
\begin{align}
\bar {\cal E}_{S V}(\rho) &= {\mathbb E}_{V\in \mu_H}\left[ {\cal E}_{S \tilde V}(\rho)\right] \nonumber\\
& =  \Lambda_S \left( \frac{\hat I_d}{d} \otimes  \tr_A (\rho)\right)  \Lambda_S^\dagger.
\label{eqn:barESV1}
\end{align}
We insert the shift operator $\Lambda_S$ with equation \ref{eqn:LambdaS0} 
   giving 
\begin{align}
\bar {\cal E}_{S V} (\rho) 
& = \frac{1}{d} \sum_{j=1}^d \ket{j}\bra{j} \otimes \sum_{j=1}^d\sum_{v,v'=1}^{N}  \bra{v} \tr_A(\rho) \ket{v'}  \nonumber \\
& \qquad \qquad  \times \ket{w(j,v)}   \bra{w(j,v')}
. \label{eqn:barESV}
\end{align}
The averaged channel can also be written in terms of the operator $\hat W_j$ in equation \ref{eqn:Wop} as 
\begin{align}
\bar {\cal E}_{SV} (\rho) &= \frac{1}{d} \sum_{j=1}^d \ket{j}\bra{j} \otimes  \hat W_j \tr_A( \rho) \hat W_j^\dagger.
\label{eqn:barESV_Wop}
\end{align}
Taking the trace of this equation over the coin space we can construct a quantum channel 
operating on the $N$-dimensional subspace 
\begin{align}
\bar {\cal E}_{BSV}(\rho_B) =  \tr_A (\bar {\cal E}_{SV} (\rho))  = \frac{1}{d} \sum_{j=1}^d \hat W_j \rho_B \hat W_j^\dagger.
\label{eqn:calEB}
\end{align}
This quantum channel describes a quantum process with a probability of $1/d$ of transition along each edge
of the graph and illustrates that the quantum channel constructed by taking the average over the 
random coin is decoherent and resembles a classical random walk on the graph with a balanced coin. 
The averaged channel resembles the decoherent open quantum channels for quantum 
walks studied by \citet{Brun_2003}. 

\subsection{A numerical illustration of the Haar-random coin quantum walk on the cycle graph}
\label{sec:illust}

\begin{figure}[htbp]\centering
 \includegraphics[width=3.3truein,trim = 10 0 0 0, clip]{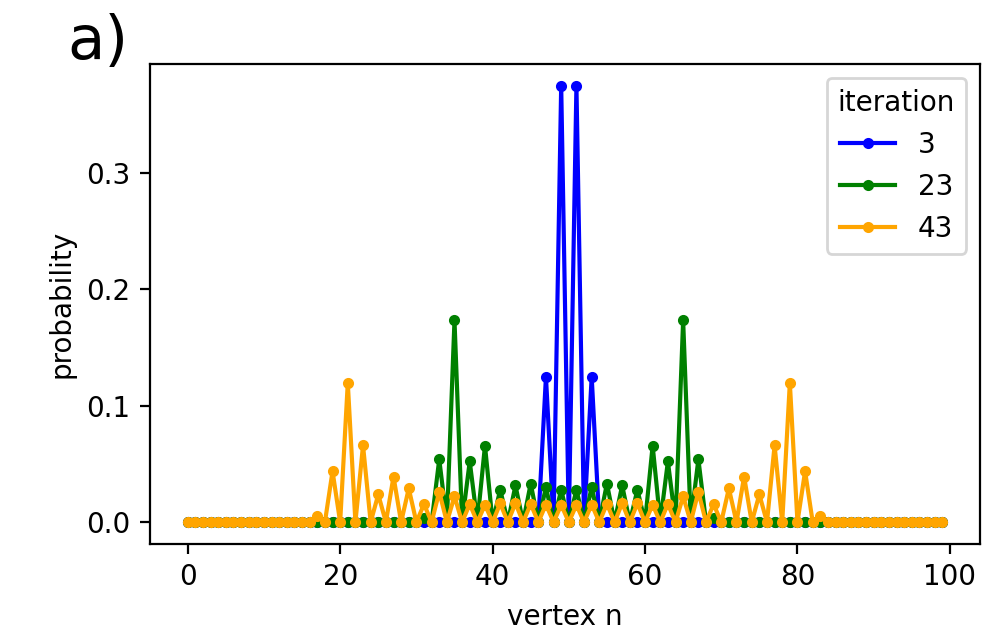}
  \includegraphics[width=3.3truein,trim = 10 0 0 0, clip]{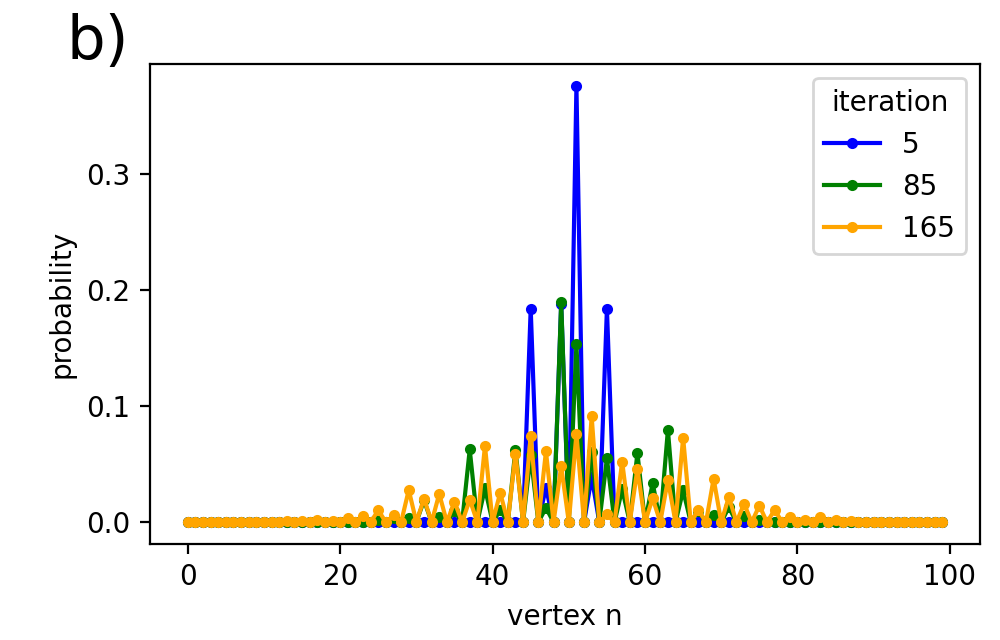}
 \caption{a) An illustration of the quantum random walk on the cycle graph with a Hadamard-coin. 
 The initial state was $\frac{1}{\sqrt{2}} ( \ket{0} + i\ket{1}) \otimes \ket{50}$
 in a $2 \times 100$ dimensional state space. Plotted are probabilities of measuring the system to 
 be in the $n$-th vertex state at iterations 3, 23 and 43. 
 b) Similar to a) except we show the associated  quantum walk with a Haar-random coin 
 at iterations 5, 85, 165. 
 This quantum walk is begun with the same initial condition as that in a).  
 The Haar-random coined quantum walk does not spread rapidly 
 like the Hadamard walk.  It spreads with a distribution 
 that is approximately Gaussian.  Decoherence caused by
 the coin causes the vertex distribution to resemble that of a 
 classical random walk on the same graph with a fair coin. 
  \label{fig:walks}
 }
\end{figure}

As previously described \citep{Aharonov_1993,Kempe_2003}, due to interference caused 
by the unitary coin operator, a quantum random walk's behavior 
differs from that of a classical random walk.    
To show that a Haar-random coin gives behavior similar to the associated classical random walk,  
we compare probabilities of position measurements for the quantum walk on the cycle 
graph with a Hadamard coin (described in section \ref{sec:Hcycle}) and that on the same graph with a Haar 
random coin (described in section \ref{sec:Haar}).  
For both cases the dimension of the quantum space is $2 \times N$ with $N=100$ and 
the system is initialized in the $\ket{\psi_0} = \frac{1}{\sqrt{2}} (\ket{0} + i \ket{1}) \otimes \ket{N/2}$ state. 
After $t$ iterations of the walk, the system is in the $\ket{\psi}^{(t)}$ state.   
In both walks the system remains in a pure state. 
In Figure \ref{fig:walks}, 
we plot the probability $p_{n,t}  = |\bra{0} \otimes \bra{n} \ket{\psi}^{(t)}  |^2 + |\bra{1} \otimes \bra{n} \ket{\psi}^{(t)} |^2$  as a function of vertex position $n \in \{0, 1, \ldots N-1\}$ at three different iteration times.  
The top panel shows a Hadamard coin quantum walk and the bottom panel shows a similar
random walk that has a Haar-random or random matrix coin. 

After a series of iterations, the distribution of walkers for the Haar-random coin 
resembles a Gaussian distribution and in that sense is similar to a classical random walk. 
In contrast,  the quantum random walk has a probability distribution for position in the $N$ state subsystem 
 that is much broader than the classical one.  The difference between the two types of walks 
 is due to interference that is present in the walk with the Hadamard coin and
 decoherence caused by the Haar-random coin in the quantum walk with 
 the Haar-random coin. 
 
\subsection{Expectation values of the iterated Markov coin random walk}

On a state $\rho$, 
we apply the random walk twice, each time with an independently chosen random matrix coin. 
With equation 
\ref{eqn:ESV_def}
\begin{align}
{\cal E}_{S \tilde V_1}( {\cal E}_{S \tilde V_2} (\rho))& =
 \Lambda_S (\hat V_1 \otimes \hat I_N) \Lambda_S (\hat V_2 \otimes \hat I_N) \rho  \nonumber \\ 
& \qquad \times ({\hat {V_2}}^{\dagger }   \otimes \hat I_N )  \Lambda^\dagger_S  (\hat V_1^\dagger \otimes \hat I_N) \Lambda_S^\dagger 
\end{align}
with $\hat V_1, \hat V_2 \in \mu_H$ both uniformly distributed with respect to the Haar measure in $U(d)$. 
As $\hat V_1 $ and $\hat V_2$ are independently drawn from the distribution $\mu_H$, 
averaging of $\hat V_1$ and $\hat V_2$ can be computed in either order;  
\begin{align}
 {\mathbb E} &_{\hat V_1, \hat V_2 \in \mu_H} \Big[{\cal E}_{S \tilde V_1}( {\cal E}_{S \tilde V_2} (\rho))\Big]
\nonumber  \\
&=  {\mathbb E}_{\hat V_1 \in \mu_H} \Big[{\cal E}_{S \tilde V_1}\left(  {\mathbb E}_{\hat V_2 \in \mu_H}\left[ {\cal E}_{S \tilde V_2} (\rho)\right]  \right) \Big]
\nonumber  \\
& =  {\mathbb E}_{\hat V_1 \in \mu_H} \Big[ \Lambda_S (\hat V_1 \otimes \hat I_N) \Lambda_S \left( \frac{\hat I_d}{d} \otimes 
\tr_A \rho \right) \Lambda_S^\dagger   (\hat V_1^\dagger \otimes \hat I_N)  \Lambda_S^\dagger \Big] \nonumber \\
& = \Lambda_S \left( \frac{\hat I_d}{d}  \otimes \tr_A \left( \Lambda_S \left( \frac{\hat I_d}{d} \otimes  \tr_A \rho  \right) \Lambda_S^\dagger \right) \right) \Lambda_S^\dagger  \nonumber \\
& = \bar {\cal E}_{SV}^{(2)} (\rho).
\end{align}
Taking the average over both $\hat V_1, \hat V_2 \in \mu_H$ is equivalent to 
two iterations of the averaged channel of equation \ref{eqn:barESV1}.
If the random walk is iterated $t$ times and during each application an independent random matrix coin is 
chosen, then the expectation value of the iterated channel is equal to the averaged channel iterated $t$ times. 
In other words 
\begin{align}
{\mathbb E}_{\hat V_1, \hat V_2, \ldots, \hat V_t \in \mu_H} \Big[ {\cal E}_{S\tilde V_1} ({\cal E}_{S\tilde V_2}( \ldots {\cal E}_{S\tilde V_t}( (\rho) ))\Big]
= \bar {\cal E}_{SV}^{(t)} (\rho) . \label{eqn:EVVV}
\end{align}

\subsection{The averaged channel is not ergodic} 
\label{sec:not_ergodic}

We consider a Haar-random coined quantum walk on a $d$-regular undirected connected and simple graph
(not necessarily a Cayley graph). 
We show here that as long as we can construct a unitary shift operator in the form 
of equation \ref{eqn:Wop}, the averaged channel (equation \ref{eqn:barESV_Wop}) of the quantum walk 
is not ergodic.   

To construct the channels discussed in this manuscript, the $d$-regular undirected, simple, 
and connected graph must allow construction of 
the index labeling index functions $w(j,v)$ described in section \ref{sec:dreg}.   
 With the edge labeling index functions on a graph with $N$ vertices 
 it is convenient to compute 
\begin{align}
\frac{1}{N} \sum_{v=1}^N \ket{w(j,v)} \bra{w(j,v)} = \frac{\hat I_N}{N}.  \label{eqn:vw1}
\end{align} 
This follows because the set $\{ w(j,v) : \forall \text{ vertices } v \}$,  
must contain all the vertices and this must be
true for all values of $j \in \{ 0, 1, \ldots, d\}$.  
We also compute 
\begin{align}
\frac{1}{N} \sum_{v,v'=1}^N  \ket{w(j,v)} \bra{w(j,v')} = \ket{y}\!\bra{y} \label{eqn:vw2}
\end{align}
with superposition state $\ket{y} \equiv  \frac{1}{\sqrt{N}} \sum_{v=1}^N \ket{v}$. 
 
A basis for operators (with respect to the Frobenius inner product) in the vertex subspace 
is the set 
of elements $\ket{v}\bra{v'}$ with $v, v' \in \{1, \ldots, N\}$. 
The averaged channel is that given by equation \ref{eqn:barESV_Wop}.  
We compute 
\begin{align}
\bar {\cal E}_{SV} \left(\tfrac{\hat I_d }{d} \otimes \ket{v}\bra{v'}\right)  =  \frac{1}{d} \sum_{j=1}^d
   \ket{j}\bra{j} \otimes  \ket{w(j,v)} \bra{w(j,v')} . \label{eqn:vvvv}
\end{align}

We consider two states 
\begin{align}
\frac{\hat I_d}{d} \otimes \sum_{v = 1}^N \ket{v}\bra{v} \nonumber \\
\frac{\hat I_d}{d} \otimes \ket{y}\bra{y} .  \label{eqn:twof}
\end{align}
When we operate on these states with the averaged channel, 
equations \ref{eqn:vw1},  \ref{eqn:vw2}  and equation \ref{eqn:vvvv}
imply that they are both eigenstates of the averaged channel with eigenvalue 1. 
Since these states have trace of 1 and are positive, they are both in the space of density operators. 
The averaged  
 channel has more than one fixed point  (in the space of density operators) 
 and consequently it is not ergodic.   We find that the averaged 
 channel of a quantum random walk with a Haar-random coin is not ergodic 
when constructed on 
any $d$-regular graph that allows ordering of edges giving unitary operators in the form of $\hat W_j$  (equation \ref{eqn:Wop}).  

The first of the states in equation \ref{eqn:twof}  is proportional to the identity 
and is completely depolarized. However because the averaged channel is not ergodic,  
iteration of the channel does not necessarily cause a density operator to approach the depolarized state. 
This implies that information about the initial density operator is retained 
even when the channel is iterated many times.  
In other words $\bar {\cal E}_{SV}^{(t)}(\rho_0) $ retains information about $\rho_0$ 
even if $t$ is large. 
 Examples discussed
in section \ref{sec:ex} of Cayley graphs on Abelian groups (including
the cycle and hypercube graphs) we find additional fixed points of the averaged channel.  

\subsection{The probability distribution of vertex measurements} 

Following \citet{Aharonov_2001} a quantum walk can be characterized with  
 the probability distribution of vertex position measurements.  
The measurement of vertex or node position is a projective measurement with 
a set of vertex measurement operators $ \{ P_v   : \  v \in \{1, \ldots, N\} \}$
with each projection operator associated with a single vertex 
\begin{align} 
 \hat P_v = \hat I_d \otimes \ket{v}\bra{v} . \label{eqn:Pv}
\end{align}   
For a specific density operator $\rho \in {\cal L} ({\mathbb C}^{dN} )$  describing a quantum state, 
we can compute the probability 
\begin{align} 
p_v(\rho) = \tr( \hat P_v \rho)   \label{eqn:measure_pv}
\end{align}  for the quantum system 
to be measured at a vertex $v$.   
 As $\sum_{v=1}^{N} \hat P_v  = \hat I_d \otimes \hat I_N$ is the identity, 
the set of nonnegative real numbers $\{ p_v :\  v \in \{1, \ldots, N \} \} $ is a normalized probability distribution. 
After $t$ iterations of a quantum walk, we can determine whether the vertex probability distribution $\{p_v(\rho) :  \forall v \} $  approaches a stationary distribution in the limit of many iterations (large $t$). 
Equivalently for initial density operator $\rho_0$  if 
\begin{align}
\lim_{k \to \infty} p_v ({\cal E}^{(t)} (\rho_0) )  = \frac{1}{N}  \label{eqn:lim}
\end{align}
for iterations of the channel ${\cal E}$, 
then the vertex probability distribution approaches a stationary distribution.

For the quantum walk with a Haar-random coin and using equation \ref{eqn:EVVV} 
\begin{align}
 {\mathbb E}_{\hat V_1, \hat V_2, \ldots, \hat V_k \in \mu_H} & \Big[ \tr \left(  \hat P_v {\cal E}_{S\tilde V_1} ({\cal E}_{S\tilde V_2}( \ldots {\cal E}_{S\tilde V_k}( (\rho) )) ) \right)\Big] \nonumber \\
&=\tr \left( \hat P_v \bar {\cal E}_{SV}^{(k)} (\rho) \right)  \nonumber \\
& = p_v \left(  \bar {\cal E}_{SV}^{(k)}(\rho) \right).
\end{align}
If the averaged channel obeys equation \ref{eqn:lim} 
then on average 
 the vertex probability distribution approaches a stationary distribution for the iterated stochastic channel. 

We show below in section \ref{sec:ex} that after iteration of the averaged channel,   
a density operator can approach one with 
a stationary and uniform vertex probability distribution, while maintaining 
correlations that would allow measurements to reveal information about the initial density operator.  
This is possible because the averaged channel is not ergodic (as discussed in section \ref{sec:not_ergodic}).

\section{Examples of averaged channels} 
\label{sec:ex}

We illustrate properties of averaged channels (taking the expectation value over the distribution 
of  coin operators) 
for the quantum walk on the cycle graph and for the walk on the hypercube. 
In the last subsection below we consider the more general case of a 
Cayley graph of a finite Abelian group. 

\subsection{Example 1: averaged channel for the quantum walk on the cycle graph}
\label{sec:acirc}

Using shift operator $\Lambda_{S, \text{cycle}} $ from equation \ref{eqn:LS_cycle} 
for the quantum walk on the cycle graph with a Haar-random coin, the averaged channel (equation \ref{eqn:barESV1} or \ref{eqn:barESV})  becomes 
\begin{align}
\bar {\cal E}_{SV,\text{cycle}}(\rho) 
 & = \tfrac{1}{2} \ket{0}\!\bra{0}
    \otimes \!\!\! \sum_{v,v'=0}^{N-1} \!\! \ket{v + 1} \!
    \bra{v'+1} \bra{v} \tr_A( \rho) \ket{v'}  \nonumber \\ 
&  +  \tfrac{1}{2} \ket{1}\! \bra{1} 
    \otimes\!\!\!\sum_{v,v'=0}^{N-1} \!\!
\ket{ v - 1} \! \bra{v'-1} \bra{v}\tr_A( \rho) \ket{v'} . \label{eqn:ESV_cycle}
\end{align}
Here we have omitted the modulo operator but it is still implied. 
This channel describes a walker that with 1/2 probability 
moves a step clockwise on the cycle graph and with 1/2 probability moves 
in the opposite direction. 
In this sense the averaged quantum walk recovers the classical random walk with a fair coin.

The Hadamard walk on the line is simpler with vertex space transformed to a Fourier basis (e.g., \citealt{Aharonov_2001}).  For the cycle graph we use the discrete Fourier transform basis 
\begin{align}
\ket{k}_F =\frac{1}{\sqrt{N}} \sum_{v=0}^{N-1} \omega^{kv} \ket{v} \qquad \text{ for }   k \in \{0, \ldots, N-1\}
\label{eqn:k_F}
\end{align} 
with 
\begin{align} \omega \equiv e^\frac{2 \pi i}{N}.\end{align}  
For calculations it is convenient to recognize that 
\begin{align}
\bra{v} \ket{k}_F = \frac{\omega^{vk}}{\sqrt{N}}.  \label{eqn:vk}
\end{align}
The shift operation is diagonal in the Fourier basis;  
\begin{align}
\sum_{v=0}^{N-1} \ket{v+1} \bra{v} = \sum_{k=0}^{N-1} \omega^{-k} \ket{k}_F\bra{k}_F. 
\end{align}
The averaged channel of equation \ref{eqn:ESV_cycle}
\begin{align}
\bar{\cal E}_{SV,\text{cycle}}(\rho)  &= \frac{1}{2} \sum_{k,k'=0}^{N-1}(  \ket{0}\bra{0}\omega^{k'-k}  + \ket{1}\bra{1}\omega^{k-k'}) \otimes \nonumber \\
& \qquad \qquad \rho_{k,k'} \ket{k}_F\bra{k'}_F. \label{eqn:ESV_cyclek}
\end{align}
where 
\begin{align}
\rho_{k,k'} = \bra{k}_F \tr_A \rho \ket{k}_F.
\end{align}

Viewing the channel as a linear superoperator, a set of eigenstates that spans  
the space of linear operators on ${\cal L} ({\mathbb C}^2 \otimes {\mathbb C}^N)$ 
is the following   
\begin{align}
\hat e_{0,kk'} & = 
\frac{1}{2} (\ket{0}\bra{0}\omega^{k'-k}  + \ket{1}\bra{1}\omega^{k-k'}) \otimes
 \ket{k}_F\bra{k'}_F  \nonumber \\
\hat e_{x,kk'} &= ( \ket{0}\bra{1} + \ket{1}\bra{0}) \otimes \ket{k}_F\bra{k'}_F \nonumber \\
\hat e_{y,kk'} & = ( \ket{0}\bra{1} - \ket{1}\bra{0}) \otimes \ket{k}_F\bra{k'}_F  \nonumber \\
\hat e_{z,kk'} & =(\ket{0}\bra{0} - \ket{1}\bra{1}) \otimes \ket{k}_F\bra{k'}_F  \label{eqn:eigs1}
 \end{align}
 for each of  $k, k' \in \{0, 1, \ldots N-1\}$. 
We have labelled the eigenstates in analogy to Pauli operators.   The eigenvectors that 
 have non-zero trace are in the form of $\hat e_{0,kk}$. 
The eigenvector/eigenvalue pairs are 
\begin{align}
\left(\hat e_{0,kk'} ,   \cos \tfrac{2 \pi (k-k')}{N} \right) ,  \left(\hat e_{x,kk'},  0\right),\
  ( \hat e_{y,kk'}, 0),\  \nonumber  \\
\text{ and } ( \hat e_{z,kk'},0)  .   \label{eqn:lamkk}
\end{align} 
We take the trace of $\hat e_{0,kk'}$ 
 \begin{align}
\tr_A( \hat e_{0,kk'} ) =  \cos \tfrac{2\pi  (k -k') }{N}  \ket{k}_F\bra{k'}_F . \label{eqn:trekk}
 \end{align}

Equations \ref{eqn:trekk} and \ref{eqn:lamkk}  imply that 
the eigenvectors $\hat e_{0,kk}$ that are diagonal in the Fourier basis 
 have eigenvalue of 1 and have trace of 1. They are fixed points of the channel; 
 $\bar {\cal E}_{SV,\text{cycle} }(\hat e_{0,kk} ) = \hat e_{0,kk}$. 
Due to the existence of multiple fixed points (one for each value of $k$)  in the space of density operators, 
the averaged channel $\bar {\cal E}_{SV, \text{cycle}}$ is not ergodic. 
The number  of linearly independent fixed points is equal to the number of vertices $N$ 
and  is remarkably large. 

The set of operators composed 
of the eigenvectors in equation \ref{eqn:eigs1},  is a linearly independent set that 
spans the space of operators in ${\cal L}({\mathbb C}^{dN})$.  However, this set is not 
 an orthogonal basis with respect to the Frobenius inner product ($\tr (\hat e_{0,kk}^\dagger \hat e_{z,kk}) \ne 0 $). 
Nevertheless any density operator can be written as a vector using this set of eigenvectors as a basis.  
Iteration of the averaged channel leaves only a sum of eigenvectors with non-zero
eigenvalues which only contains a sum of $\hat e_{0,kk'}$ terms. 
The set of eigenvectors $\{ \hat e_{0,kk'}  : \ \forall k,k' \} $ are orthogonal with respect to the 
 Frobenius inner product; $\tr (\hat e_{0,jj'}^\dagger \hat e_{0,kk'} ) =\tfrac{1}{2} \delta_{jk} \delta_{j'k'}$. 

The magnitude of the eigenvalues of $\hat e_{0,kk'}$  in equation \ref{eqn:lamkk} $|\lambda_{0,kk'}| \le 1$, as expected for eigenvalues of a quantum channel. 
For eigenvalues satisfying $|\lambda_{0,kk'}| < 1$,  iteration of the channel would cause a reduction 
in the amplitude of that particular eigenvector, and in the limit of many iterations 
only those with $|\lambda_{0,kk'}| = 1$ remain. 
Based on equation \ref{eqn:eigs1}, the only eigenvectors with eigenvalues of magnitude 1 are 
\begin{align}
\hat e_{0,kk}& = \tfrac{\hat I}{2} \otimes \ket{k}_F\bra{k}_F \nonumber \\
\hat e_{0,k,k+N/2} &= \tfrac{\hat I}{2} \otimes \ket{k}_F \bra{k + \tfrac{N}{2}}_F,  \text{ if } N \text{ even}, 
\end{align}
where we omitted a global phase in the second expression. 
The eigenvalue of $\hat e_{0,kk} $ is 1 and  
the eigenvalue of $\hat e_{0,k,k+N/2}$ is $\lambda_{0,k,k+N/2} = -1$. 

For an initial state $\rho_0$, 
we compute the coefficients for each $k$ 
\begin{align} a_k &= 2\tr (\rho_0 \hat e_{0,kk}) \nonumber \\
b_{k,k+N/2} &= 2\tr (\rho_0 \hat e_{0,k,k+{N/}{2}})   \nonumber \\
b_{k+N/2,k} &=2 \tr (\rho_0 \hat e_{0,k+{N/}{2},k}) 
 \label{eqn:abk}.
\end{align}
The $b$ coefficients are only relevant for $N$ even.  
Because the density operator is Hermitian, $b_{k,k+N/2} = b_{k+N/2,k}^* $ 
and because it has a trace of 1, $\sum_{k=0}^{N-1} a_k = 1$. 
After a larger number $t$ of iterations the resulting density operator 
\begin{align}
\rho^{(t)} &=  \bar {\cal E}_{SV, \text{cycle}}^{(t)} (\rho_0) \nonumber \\
&  \sim 
 \sum_{k=0}^{N-1} \Big[ a_k \hat e_{0,kk} + b_{k ,k+N/2} \hat e_{0,k,k+N/2} (-1)^t  \nonumber \\
& \quad \ \ + b_{k+N/2,k} \hat e_{0,k+N/2,k} (-1)^t\Big]
 \label{eqn:large_t},
 \end{align}
 where I have used the eigenvalues of the eigenvectors and neglected all decaying eigenstates. 
 This operator encodes information that can be retrieved from measurements 
 that is retained upon iteration of the channel. 

Using Equation \ref{eqn:vk}
\begin{align}
\bra{v} \hat e_{0,kk} \ket{v} &= \tfrac{1}{N} \nonumber \\
\bra{v} \hat e_{0,k,k+N/2} \ket{v} & = \tfrac{(-1)^v}{N} .
\end{align}
Using equation \ref{eqn:measure_pv} and equation \ref{eqn:vk} the probably of measuring 
the system to be at vertex $v$  after a large number of iterations of the averaged channel is 
\begin{align}
p_v(\rho^{(t)}) \sim \tfrac{1}{N} \left( 1 - \sum_{k=0}^{N-1}  (\text{Re}[b_{k,k+N/2} ] (-1)^{t+v} \right).
\label{eqn:pvt0}
\end{align}
If $N$ is odd, the probability $p_v=1/N$ is independent of initial conditions and corresponds to a uniform vertex distribution, recovering the limiting distribution expected for the classical random walk. 
In this case vertex measurements do not reveal any information about the initial conditions. 
If $N$ is even, then the probabilities are sensitive to $b_{k,k+N/2}$ which depends upon 
the initial conditions through equation \ref{eqn:abk}. In this case vertex measurements 
can reveal information about the initial conditions. 
 
What types of initial states would have an off-diagonal $k,k+N/2$ component? Assuming $N$ even, 
consider a pure initial state that is a superposition state 
of the  $k$ and $k+N/2$ Fourier components that depend on the angle $\theta \in [0, \pi]$, 
and a relative phase $\phi \in [0, 2 \pi]$, 
\begin{align}
\ket{\psi_{0a}}= \ket{c} \otimes \left( \cos \tfrac{\theta}{2} \ket{k}_F +e^{i \phi} \sin \tfrac{\theta}{2 }\ket{k + N/2}_F \right),
\end{align} 
and with $\ket{c}$ is any normalized state in the coin subspace. 
We use this superposition state to construct a initial density operator 
\begin{align}
\rho_{0a} & = \ket{\psi_{0a}} \bra{\psi_{0a}} \nonumber \\
 &=\ket{c}\!\bra{c} \otimes \big[  \cos^2\! \tfrac{\theta}{2}  \ket{k}_F\!\bra{k}_F 
+ \sin^2\! \tfrac{\theta}{2} \ket{k+\tfrac{N}{2}}_{\! F}\!\bra{k+\tfrac{N}{2}}_F\nonumber \\
& \ \ +  \tfrac{1}{2} \sin \theta \left(e^{-i\phi} \ket{k}_F \bra{k+ \tfrac{N}{2}}_{F}  + e^{i\phi} \ket{k + \tfrac{N}{2}}_F \bra{k}_F  \right)
\big].
\end{align}

Using equation \ref{eqn:abk}, 
we compute coefficients for this initial density operator 
\begin{align}
a_k &= \cos^2\tfrac{\theta}{2}, \quad a_{k+N/2} = \sin^2 \tfrac{\theta}{2}, \nonumber \\   
b_{k,k+N/2} &= \tfrac{e^{-i\phi} }{2} \sin \theta, \ \  b_{k+N/2,k} =  \tfrac{e^{i\phi} }{2} \sin \theta.
\end{align}
After iteration of the channel $t$ times,  equation \ref{eqn:large_t} gives a density operator 
\begin{align}
\rho^{(t)}_a &  =  \bar {\cal E}_{SV, \text{cycle}}^{(t)} (\rho_{0a}) \nonumber \\
  & = \frac{\hat I}{2} \otimes 
\cos^2 \tfrac{\theta}{2}  \ket{k}_F\bra{k}_F 
+ \sin^2 \tfrac{\theta}{2} \ket{k+\tfrac{N}{2}}_F\bra{k+\tfrac{N}{2}}_F \nonumber \\
& \ \ + \tfrac{ ( -1)^{t}}{2} \sin \theta \left(e^{-i\phi} \ket{k}_F \bra{k\!+\! \tfrac{N}{2}}_F  + e^{i\phi}\ket{k\!+\! \tfrac{N}{2} }_{\!F} \! \bra{k}_F  \right ).
\end{align}

Note the initial state $\ket{\psi_{0a}}$ is a pure state and the quantum walk with the Haar random 
coin is unitary giving pure states during each iteration.   
The averaged channel is an average  
over of the possible randomly chosen coin operators.  The density operator $\rho^{(t)}_a$ 
is a mixture of pure states taking into account the averaging. 

Using equation \ref{eqn:pvt0}  the probability of measuring the system with density operator $\rho^{(t)}_a$ to 
be at vertex $v$ after $t$ iterations of the averaged channel 
\begin{align}
p_v(\rho^{(t)}_a) &= 
\frac{1}{N} + \frac{(-1)^{t+v} }{N} \sin \theta \cos \phi.
\end{align}
The distribution of vertex position measurements would not approach 
a uniform distribution and a series of measurements could reveal information about the initial state. 
A series of measurements after many iterations of the averaged channel would allow one 
to measure the product  $\sin \theta\cos \phi$.   Despite the fact that the coin operator 
is completely depolarizing, it is possible to learn something about the initial state, even after a large 
number of iterations (but only if $N$ is even).   
With vertex position measurements alone 
you could not learn everything about the state 
as you would not be able to constrain $\theta$ independent of $\phi$ and 
nothing is learned about the initial coin state $\ket{c}$. 
With other types of measurements (such as those in the Fourier basis or that involve correlations between vertices) it would be possible to constrain both $\phi$ and 
 $\theta$ as that information is retained in the density operator.

We now consider the setting where $N$ is odd and all off diagonal elements 
of the density operator decay upon iteration of the averaged channel. 
We have established that vertex position measurements approach a uniform distribution, 
but is it possible to design other types of measurements 
that can reveal information about the initial state, even after many iterations of the channel? 
Inspection of the density operator after many iterations (equation \ref{eqn:large_t}) 
shows that only the information about the initial Fourier amplitudes $a_k$ is retained.  
In this sense the channel is dephasing. 
While the amplitudes cannot be measured with vertex position measurements, 
they can be measured from measurements in the Fourier basis 
or from measurements of correlations between vertices. 
For example, in analogy to a Pauli $X$ operator we can carry out a measurement that 
measures correlations between vertex positions  using 
the Hermitian operator 
\begin{align}
\hat P_{vv'} &= \tfrac{1}{2} \hat I_d \otimes( \ket{v'}\bra{v} + \ket{v'}\bra{v})   \label{eqn:Pvv}
.
\end{align}
The expectation values for $\rho^{(t)}$ of equation \ref{eqn:large_t} (but without $b$ coefficients)
\begin{align}
\tr (\rho^{(t)} \hat P_{vv'}) &=   \tfrac{1}{N} \sum_k a_k \cos\left( \tfrac{2 \pi (v'-v)k}{N} \right) .
\end{align}
For $v \ne v'$,  these measurements are sensitive to the initial Fourier amplitudes. 

\subsection{Example 2: averaged channel for the walk on the hypercube graph}
\label{sec:ahyp}

Using shift operator $\Lambda_{S, \text{hypercube}} $ from equation \ref{eqn:LS_hypercube} for the quantum 
walk on the hypercube with Haar-random coin, 
the averaged channel (equation \ref{eqn:barESV}) becomes
\begin{align}
\bar {\cal E}_{SV,\text{hypercube}}(\rho) & =\frac{1}{d} \sum_{j=1}^d \ket{j}\bra{j} \otimes 
\hat X_j  \rho_B \hat X_j .
\end{align}
This channel describes a walker with $1/d$ probability 
 moving along each of the $d$ edges on the hypercube.  
In this sense the averaged quantum walk recovers the classical walk 
on the hypercube with a fair $d$-sided dice.  

From the averaged quantum channel, 
we can construct a quantum channel that operates on the density operator in the $N$-dimensional subspace
with equation \ref{eqn:calEB}, 
\begin{align}
\bar {\cal E}_{BSV} (\rho_B)_\text{hypercube} &= \tr_A \left( \bar{\cal E}_{SV,\text{hypercube}}\left( \tfrac{\hat I_d}{d} \otimes \rho_B\right)\right) \nonumber \\
& =  \tfrac{1}{d} \sum_{j=1}^d \hat X_j \rho_B \hat X_j. \label{eqn:calEB_hyp}
\end{align}
In this form we can see that the set of operators 
$\{ \hat X_j/\sqrt{d}  \ :   j \in \{1, \ldots, d \} \} $ (proportional to the Pauli $X$ operators) are a complete set of Kraus operators for an operator-sum decomposition of the averaged quantum channel
restricted to the vertex subspace. 

The relevant discrete Fourier transform would be that associated with the Abelian group $ (\oplus {\mathbb Z}_2)^{d}$. 
We use a basis $\ket{k}_F$ with $k = k_1 k_2 \ldots k_d  \in \{0,1\}^{\otimes d}$ written as a binary string 
\begin{align}
\ket{k}_F = \frac{1}{2^{d/2}} \sum_{{v} \in \{0,1\}^{\otimes d} } (-1)^{{ k} \cdot { v}} \ket{v} 
\end{align}
where ${ k} \cdot { v} = \sum_{m=1}^d k_m v_m$ is the dot product of the two binary digit strings. 
It is convenient to compute 
\begin{align}
\bra{v}\ket{k}_F = \frac{1}{2^{d/2} } (-1)^{ v \cdot k} . \end{align} 

We transform the Pauli operator $\hat X_j$ (operating on the $j$-th qubit) into the Fourier basis   
\begin{align}
\hat X_j = \sum_{k \in \{0, 1\}^{\otimes d}} \ket{k}_F\bra{k}_F (-1)^{k_j}. \label{eqn:Xjhy}
\end{align}

Using $\rho_{kk'} = \bra{k}_F \tr_A( \rho ) \ket{k'}_F$,
and equation \ref{eqn:Xjhy}, the averaged channel  of equation 
\ref{eqn:calEB_hyp} in the Fourier basis is 
\begin{align}
\bar {\cal E}_{SV}(\rho)_\text{hypercube} & = \frac{1}{d} \sum_{j=1}^d  \ket{j}\bra{j} \otimes \!\!\!
\sum_{k,k' \in \{0, 1\}^{\otimes d}}  \ket{k}_F\bra{k'}_F \nonumber \\
& \qquad \times \rho_{kk'} (-1)^{k_j+ k'_j}. \label{eqn:kk_hyp}
\end{align}

As done in the previous section for the circle graph (section \ref{sec:acirc}), 
we construct a set of eigenstates of the channel's superoperator 
\begin{align}
\hat e_{0,kk'}& = \tfrac{1}{d}  \left( \sum_{j=1}^d  \ket{j}\bra{j} (-1)^{k_j+ k'_j} \right) 
 \otimes  \ket{k}_F\bra{k'}_F \nonumber \\
 \hat e_{jj',kk'} & = \ket{j}\bra{j'} \otimes  \ket{k}_F \bra{k'}_F \text { for } j \ne j' \nonumber \\
 \hat e_{j,kk'} & = (\ket{j}\bra{j} - \ket{j+1} \bra{j+1}) \otimes \ket{k}_F \bra{k'}_F \nonumber \\ 
 & \qquad \qquad \text { for } j < d 
  . \label{eqn:xkk_hyp}
 \end{align}
The eigenvalues of the eigenvectors (listed in equation \ref{eqn:xkk_hyp}) 
are zero except for those in the form of $\hat e_{0,kk'}$, 
 in which case their eigenvalues are 
\begin{align}
\lambda_{0,kk'}  = \frac{1}{d} \sum_{m=1}^d (-1)^{k_m+ k'_m} . \label{eqn:lam0kk_hyp}
\end{align}
Here $k_m, k_m'$ are the $m$-th digits of labelling strings $k,k'$.

The eigenvectors listed in equation \ref{eqn:xkk_hyp}  are 
 not orthogonal with respect to the Frobenius inner product 
but they are linearly independent and  span the space of operators in ${\cal L} ({\mathcal C}^{dN})$. 
The eigenvectors in the set $\{ \hat e_{0,kk'}  : \  k,k' \in \{0, 1\}^{\otimes d} \} $ are orthogonal with respect to the Frobenius
inner product as $\tr ( \hat e_{0,kk'}^\dagger  \hat e_{0,mm'}) = \tfrac{1}{d}\delta_{km} \delta_{k'm'}$. 
We can write an initial density operator as a vector sum of the eigenvectors listed in equation \ref{eqn:xkk_hyp}. 
After many iterations, only terms dependent on eigenvectors that have 
eigenvalues $|\lambda_{0,kk'} | =1$ would remain. 

With $k=k'$, equation \ref{eqn:lam0kk_hyp} shows that eigenvector $\hat e_{0,kk}$ has 
 eigenvalue $\lambda_{0,kk}= 1$.  
As the trace of $\hat e_{0,kk} = 1$ for any  
$k \in \{0, 1\}^{\otimes d}$, 
the eigenvectors $\hat e_{0,kk}$ are fixed points in the space of density operators of
the averaged channel $\bar {\cal E}_{SV,\text{hypercube}}$. 
As there are multiple density operator fixed points, the averaged channel is not ergodic. 

The only eigenstates that have eigenvalue of 1 are $\hat e_{0,kk}$. 
We find that  
 $\lambda_{0,kk'}  = -1 $ only if each digit of $k$ is the logical not of each digit of $k'$. 
 We denote these as $\hat e_{0,k \bar k}$. 
 As we did in the example of the cycle graph (section \ref{sec:acirc}), 
 we can consider a pure superposition state that gives a component of $\hat e_{0,k\bar k}$. 
 We denote $\ket{\tilde 0}_F $ and $\ket{\tilde 1}_F$ to be the state in the Fourier basis 
 with all digits of $k$ equal to 0, and 1 respectively. 
A normalized initial superposition state dependent upon angles $\theta$  and $\phi$ 
 \begin{align}
 \ket{\psi_{0c}} = \ket{c} \otimes( \cos \tfrac{\theta}{2} \ket{\tilde 0}_F  
  + e^{i\phi} \sin \tfrac{\theta}{2} \ket{\tilde 1}_F)
 \end{align}
 for any state $\ket{c}$ in the coin space. 
 The initial density operator for this state 
 \begin{align}
 \rho_{0c} &=\ket{\psi_{0c}} \bra{\psi_{0c}} \nonumber \\
 &= \ket{c}\bra{c} \otimes \Big( \cos^2 \tfrac{\theta}{2}  \ket{\tilde 0}_F\bra{\tilde 0}_F 
 +  \sin^2 \tfrac{\theta}{2}  \ket{\tilde 1}_F\bra{\tilde 1}_F  \nonumber \\
 & + \tfrac{1}{2} \sin \theta \left( e^{-i\phi } \ket{\tilde 0}_F\bra{\tilde 1}_F  +  e^{i\phi } \ket{\tilde 1}_F\bra{\tilde 0}_F\right) \Big) .
 \label{eqn:rho0_hyp}
 \end{align}
 We compute the strength of the coefficients for a vector expansion in terms of eigenstates 
 \begin{align}
 a_0 &= d\tr ( \rho_{0c} \hat e_{0,\tilde 0\tilde 0} ) =  \cos^2 \tfrac{\theta}{2} \nonumber \\
 a_1& = d\tr ( \rho_{0c} \hat e_{0,\tilde 1\tilde 1} ) =  \sin^2 \tfrac{\theta}{2}  \nonumber  \\
 b_{ 0 1} & =  d\tr ( \rho_{0c} \hat e_{0,\tilde 0\tilde 1} ) = -\tfrac{e^{i\phi}}{2} \sin \theta  \nonumber  \\
 b_{ 10} & =  d\tr ( \rho_{0c} \hat e_{0,\tilde 1\tilde 0} ) = -\tfrac{e^{-i\phi}}{2} \sin \theta .
 \end{align}
   
We compute the probability of measuring the system at vertex $v$ for the channel's eigenvectors 
using  equation \ref{eqn:xkk_hyp}
 \begin{align}
 p_v(\hat e_{0,kk'}) &  = \tr (\hat P_v \hat e_{0,kk'} ) \nonumber \\
 & = \frac{1}{d} \sum_{m=1}^d (-1)^{k_m + k_m'} \bra{v} \ket{k}_F \bra{k'}_F \ket{v} \nonumber \\
 & = \frac{2^{-d}}{d} \sum_{m=1}^d (-1)^{k_m + k_m'} (-1)^{v\cdot (k + k')} .
 \end{align}
 If $k=k'$
 \begin{align}
  p_v(\hat e_{0,kk})  =  \tfrac{1}{2^d},
 \end{align}
 and the probability is independent of the vertex $v$.  
 However if $k$ and $k'$ differ in all their digits ($k' =  \bar k $) then 
 \begin{align}
 p_v(\hat e_{0,k\bar  k }) 
 & = - \tfrac{1}{2^d} (-1)^{ \text{parity} (v)}, 
 \end{align}
 where the parity function parity$(v) =1$ if the sum of the digits of $v$ is odd and is 0
 otherwise. 
 
For initial state $\rho_{0c}$ given in equation \ref{eqn:rho0_hyp} and iteration number $t$ large 
 \begin{align}
 \rho^{(t)}_c &= \bar {\cal E}_{SV,\text{hypercube}}^{(t)} (\rho_0)\nonumber \\
 & \sim \cos^2 \tfrac{\theta}{2} \hat e_{0,\tilde 0\tilde0} + \sin^2 \tfrac{\theta}{2} \hat e_{0,\tilde1\tilde1} \nonumber \\
 & \qquad - (-1)^t \tfrac{\sin \theta}{2}  ( e^{-i \phi} \hat e_{0,\tilde 0 \tilde1} + e^{i\phi} \hat e_{0,\tilde1\tilde0}). 
 \end{align}
Note that the information describing the initial superposition state (in terms of angles $ \theta, \phi$) 
is present in the density operator which implies that it is measurable. 
A measurement of the vertex position gives probability 
 \begin{align}
 p_v (\rho^{(t)}_c) & =  \tr ( \rho^{(t)}_c \hat P_v) \nonumber \\
 & \sim \frac{1}{2^d} \left( 1 +(-1)^t \sin \theta \cos\phi (-1)^{\text{parity}(v)} \right) .
 \end{align}
The measurement probability depends on the angles $\theta,\phi$ and the parity of $v$. 
From measurements of different parity vertex positions it would
be possible to determine $\sin \theta \cos \phi$ giving a constraint on the angles.  
With additional types of measurements, such as in the Fourier basis or with correlated 
vertex position measurements,  
it would be possible to constrain both angles as the density operator contains both 
phase and amplitude information about the initial superposition state. 

The averaged quantum channel corresponding to the quantum walk with a depolarized
coin on the hypercube exhibits phenomena similar to the average quantum channel 
on the cycle graph that was discussed in section \ref{sec:acirc}. 
However for the hypercube graph, the averaged channel always has traceless off-diagonal eigenstates 
with eigenvalue -1 and that do not decay upon iteration of the channel. 
The off-diagonal eigenstates are present in an initial density operator if the initial state is a superposition 
of states $\ket{k}_F$ and $\ket{\bar k}_F$ in the Fourier basis, for any $d$ length  
bit string $k$.   Measurements can be devised that would reveal information 
about the initial superposition even after many iterations of the the quantum channel. 
Measurement  of the vertex positions only approaches a uniform distribution if 
the initial density operator does not contain these non-decaying off-diagonal terms.

\subsection{Example 3: A quantum walk on an undirected Cayley graph of an Abelian group and 
its averaged channel}
\label{sec:Cayley}

The examples of the cycle and hypercube graphs are in the class of undirected 
Cayley graphs of Abelian groups. 
In this section we add a third example that is a more general case of a quantum walk 
on an undirected Cayley graph of a finite Abelian group. 

According to the fundamental theorem of finite Abelian groups,  
any finite Abelian group can written as a direct sum of cyclic groups.   
We denote the set of group generators of group $G$ as the set $H$.    
Any element of $g \in G$ in the Abelian group $G$ 
can be written as a product of powers of the group's generators 
$g =  \prod_{j=1}^{|H|} s_j^{v_j}$ for group generator $s_j \in H$, and 
where each power $v_j \in \{0, 1, \ldots, |s_j| -1 \}$. 
With an order chosen for the generators, we can use a string $v$ 
 to specify group elements or equivalently vertices of the Cayley graph. 
The string $v$ has $|H|$ digits and each digit is an integer that 
 takes values  $v_j \in \{0, 1, \ldots, |s_j|-1 \}$.   With this notation in mind, 
the associated quantum state in the vertex subspace is $\ket{v}$ and the dimension 
of the vertex quantum subspace $N=|G| = \prod_{j=1}^{|H|} |s_j|$ is the product 
of the orders of the group's generators. 

A quantum walk can be constructed from a $d$-regular graph 
that is formed from a group $G$ using a Cayley graph \citep{Aharonov_2001,LopezAcevedo_2006}.
A Cayley graph $\Gamma(G,S)$ is generated from a group $G$ and a set $S$ of group elements. 
Each element of the group is associated with a graph vertex, and the set $S$ specifies 
the edges of the graph.  By convention, the set $S$ does not include the identity element of the group. 
The graph's edges are vertex pairs $(g,sg)$ where $g\in G$ is an element of the group,  
$s\in S$ and $sg$ denotes group multiplication of $s$ and $g$.  
If the set $S$ satisfies $s, s^{-1} \in S \  \forall s \in S$,  then the Cayley graph
is undirected. 
The number of edges per vertex ($d$, the degree) of a Cayley graph depends on the 
number of elements in the edge generating set $S$.

The cycle graph is an example of the Cayley graph $\Gamma({\mathbb Z}_N, S)$ of the cyclic group $ G ={\mathbb Z}_N$ with set $S = \{ s, s^{-1} \}$ where $s$ is an element 
that generates the group.  
The set of group generators $H= \{ s\}$, but as long as $N>1$, the
number of elements is $S$ is two because the set $S$ contains both 
the group generator and its inverse. In this case $d = |S| = 2 |H|$. 

The hypercube is also an example of the Cayley graph 
$\Gamma((\oplus {\mathbb Z}_2 )^d,S)$ of the group $G=(\oplus {\mathbb Z}_2 )^d$, 
a direct sum of $d$ copies of ${\mathbb Z}_2$ (the Abelian group with 2 elements). 
There are $d$ generators of the group in the set $S$ and every generator is its own inverse.  
In this case $S = H$ the set of generators of the group and the degree of the graph $d = |S| = |H|$. 

So that the set $S$ contains its own inverse (giving an undirected graph) 
and so that it generates the whole group $G$ 
(giving a connected graph) we require that the set $S$ contains the generators of each cyclic 
group of its decomposition into cyclic groups and their inverses. 

The set $S$ is used to generate graph edges and the operators for edge transitions 
(equation \ref{eqn:Wop}), but the Fourier transform and its associated basis 
only depend on the number of group generators.  
 
Suppose the set $H$ of generators of the group has a number $a_s$ generators of order 2. 
We order the elements of the edege set $S$ beginning with the generators of order 2, 
\begin{align}
S &\equiv \{ s_1, s_2, \ldots, s_{a_s},  \nonumber \\
   & \quad \ \  s_{a_s+1}, 
     s_{a_s+1}^{-1}, s_{a_s + 2}, s_{a_s+2}^{-1} \ldots, s_{|H|}, s_{|H|}^{-1} \} .\label{eqn:orderedL}
\end{align} 
This length of this list is equal to the degree 
$d$ of each vertex of the Cayley graph 
\begin{align} 
d =   |S| = 2 |H| - a_s \label{eqn:dCayley}. 
\end{align} 

For the Cayley graph generated by a finite Abelian group $G$, the ordering 
of the list in equation \ref{eqn:orderedL} 
based on the generators gives an ordering 
for edges and ensures that the operators 
$\hat W_m \in {\cal L} ({\mathbb C}^{|G|} ) $ (of equation \ref{eqn:Wop} and for $m\in  \{ 1, \ldots, d\}$) are unitary.  
Specifically the unitary operators use for constructing a shift operator 
\begin{align}
\hat W_m =\sum_{g\in G}  \ket{r_m g } \bra{g} \qquad \text{ for } r_m \in S,  
\end{align}
where $r_m$ represents the generator or inverse generator in the ordered list $S$. 
The shift operator used to construct a quantum walk is the same as we discussed previously 
(equation \ref{eqn:LambdaS0});
\begin{align}
\Lambda_S = \sum_{m=1}^d \ket{m}\bra{m} \otimes \hat W_m, 
\end{align}
where $\Lambda_s \in {\cal L} ( {\mathbb C}^{dN})$ with degree 
$d$ given in equation \ref{eqn:dCayley} and $N = |G|$. 

For a finite Abelian group $G$, a discrete Fourier transform \citep{Audrey_1999} can be defined using 
the group $\hat G$ of its multiplicative characters (which are the 1-dimensional irreducible representations). 
A character $\chi_k() : G \to {\mathbb C}$ (labelled by index $k$) 
generates a complex number $\chi_k(g)$ that is a root of unity  
 for each element $g$ in the group.  Multiplication of the characters obey group multiplication; 
 $\chi_k (g) \chi_k(h) = \chi_k(gh)$ for $g,h \in G$.   
 For a finite Abelian group, the group of characters contains 
 the same number of elements as the group itself. 
A discrete Fourier transform and an orthonormal Fourier 
basis is constructed from the characters with 
\begin{align}
\ket{k}_F &= \frac{1}{\sqrt{|G|}} \sum_{g \in G} \chi_k(g) \ket{g}. 
\end{align} 

For each generator $s_j \in H$ of the group we construct a complex root of unity 
\begin{align}
\omega_j  = e^\frac{2 \pi i}{|s_j|}. \label{eqn:om_j}
\end{align}
The discrete Fourier transform of a cyclic group is particularly straightforward as 
it is in the form of equation \ref{eqn:k_F}. 
A Fourier basis for a finite Abelian group can be described as a product of discrete Fourier transforms in 
the form of equation \ref{eqn:k_F}.  
Each discrete Fourier transform in the product is associated 
with a generator $s_j\in H$ (the generating set for the group) and depends on a phase factor $\omega_j$ 
as defined in equation \ref{eqn:om_j}. 

The associated Fourier vector is likewise a string $k$ where 
each digit $k_j \in \{0, 1, \ldots, |s_j| -1 \}$. 
We define a Fourier basis 
\begin{align}
\ket{k}_F &= \frac{1}{\sqrt{ |G|}}  \sum_{v\in V} \left( \prod_{j=1}^{|H|} \omega_j^{v_j k_j} \right )\ket{v} . 
\end{align}
It is convenient to compute 
\begin{align}
\bra{v}\ket{k}_F = \frac{1}{\sqrt{ |G|}}  \prod_{j=1}^{|H|} \omega_j^{v_j k_j}.
\end{align}
Here $v_j, k_j$ are the $j$-th digits of the labelling strings $v,k$. 

To find the operators $\hat W_m $ for edge transitions 
(defined in equation \ref{eqn:Wop}) in the Fourier basis, we compute 
\begin{align}
\bra{k}_F \hat W_m \ket{k'}_F  
\end{align}
for each of the elements of the ordered list $S$ of equation \ref{eqn:orderedL}  
with index $m$ specifying the specific element of the set $S$. 
We find that 
\begin{align}
\hat W_m &= \sum_k  \omega_{m}^{k_{j(m)} (-1)^{x_m}} \ket{k}_F\bra{k}_F \label{eqn:Wm}
\end{align} 
with 
\begin{align}
j(m) & = 
 \begin{cases} 
 m & \text{for } 1< m \le a_s \\
  \text{int} [(m-a_s)/2]  & \text{for } a_s < m \le d
 \end{cases}  \nonumber \\
  x_m &=
 \begin{cases} 
 0 & \text{for }1< m \le a_s \\
  (m-a_s) \text{ mod } 2 & \text{for } a_s < m \le d 
 \end{cases} \nonumber \\
 \omega_{m} & = e^\frac{2 \pi i}{|s_m|}. \label{eqn:Wm2}
\end{align}
Here $k_{j(m)}$ is the $j$-th digit in $k$ written as a string and $s_m$ is the $m$-th group 
element in the ordered list of equation \ref{eqn:orderedL}. 
Here $x_m$ serves to identity the inverse elements for elements in $S$ that 
have order greater than 2. 
The sum over $k$ is over all Fourier basis elements. 
Equation \ref{eqn:Wm} shows that 
in the Fourier basis, the unitary operators $\hat W_m \in {\cal L}({\mathbb C}^N)$ are diagonal.

As before, we construct a unitary quantum random walk with a Haar-random coin.   
The coin space has dimension $d=|S|$.  
The averaged quantum channel $\bar{\cal E}_{SV}$
is again given via equation \ref{eqn:barESV_Wop} and corresponds to an open 
quantum channel with a depolarizing coin operator.  
We transform the channel so that the vertex space is in the Fourier basis, giving 
\begin{align}
\bar {\cal E}_{SV} (\rho) &= 
\frac{1}{d} \sum_{m=1}^d \ket{m}\bra{m} \otimes 
 \hat W_m \tr_A( \rho) \hat W_m^\dagger \nonumber \\
& = \frac{1}{d} \sum_{m=1}^d \ket{m}\bra{m} \otimes 
\sum_{kk'}  \omega_m^{k_{j(m)} (-1)^{x_m}} \rho_{kk'} \ket{k}_F\bra{k'}_F \label{eqn:CESV}
\end{align}
with $\rho_{kk'} = \bra{k}_F \tr_A (\rho) \ket{k'}$ and notation given 
in equations \ref{eqn:Wm} and \ref{eqn:Wm2}. 
The sum in equation \ref{eqn:CESV} is over all strings $k, k'$ specifying states in the Fourier basis. 

The only eigenvectors of the averaged channel that don't have eigenvalue of zero are in the class 
\begin{align}
\hat e_{0,kk'}& = \frac{1}{d} \sum_{m=1}^{d}\ket{m}\bra{m} \otimes
 \omega_m^{(k_{j(m)} - k'_{j(m)})(-1)^{x_m}} \ket{k}_F\bra{k'}_F  .\label{eqn:ekkk}
 \end{align}
Here $k, k'$ can be that of any state in the Fourier basis. 
Recall these are strings with number of digits equal to $|H|$, the size of the 
generating set, and with digit possible values dependent on the order of the corresponding group 
generator element. 
The associated eigenvalues for the eigenvectors of the channel are
\begin{align}
\lambda_{0,kk'} = \frac{1}{|H|}  \sum_{j=1}^{|H|} \cos \frac{2 \pi (k_j - k_j') }{|s_j|}. \label{eqn:eigs_Cayley}
\end{align}
Eigenvectors of the averaged channel that are not in the form of equation \ref{eqn:ekkk} have eigenvalue of zero. 

As was true for the eigenvectors of the averaged channel on the cycle and hypercube graphs, 
the eigenvalues of the averaged channel for a Cayley graph 
(equation \ref{eqn:eigs_Cayley}) are 1 only if $k=k'$. 
In this case the associated eigenstate $\hat e_{0,kk}$ has trace of 1 so it is in the space of density operators. 
For each $k$ the eigenstate $\hat e_{0,kk}$  gives a fixed point of the channel, 
hence the averaged channel is not ergodic.   Again, the number of linearly independent fixed points 
is equal to the number of vertices or $|G|$  and is remarkably large. 

As we did in sections \ref{sec:acirc},  \ref{sec:ahyp}, we can write a density 
operator as a vector (in the space of linear operators) with a sum of its eigenvectors.  The only components 
that remain after many iterations of the channel are those with eigenvalue of 1 or $-1$. 
Following the arguments presented in the previous sections, a measurement of 
vertex position can reveal whether or not the initial state contains a component 
of an eigenstate with eigenvalue $-1$. 

To have an eigenvalue that has magnitude of $-1$, equation \ref{eqn:eigs_Cayley} implies that all 
generators must have even order.  
 For eigenvalue $\lambda_{0,kk'}=-1$ the following must be satisfied 
\begin{align}
k_j' = k_j+ |s_j|/2\ \text{ mod } |s_j|  \quad \text { for } j \in \{1, \ldots,|H| \} \label{eqn:condj}.
\end{align}
Here $k_j$ and $k_j'$ are the $j$-th digits of $k,k'$ 
corresponding to the $j$-th generator $s_j$ with order $|s_j|$, 
and the condition of \ref{eqn:condj} must be satisfied for all digits. 
As we showed in sections \ref{sec:acirc} and \ref{sec:ahyp}, with a superposition state, we can construct 
a density operator that has an off diagonal component that would not decay after iteration 
of the channel.  Furthermore, measurements of the vertex position would be able to determine 
whether the initial density operator contained these off diagonal states, even after the channel is iterated. 

What if there are no eigenstates in the form of $\hat e_{0,kk'}$ in equation \ref{eqn:ekkk} that have eigenvalue 
of $-1$? In this case all eigenstates that have eigenvalue with magnitude 1 are diagonal in the Fourier basis. 
For an initial density operator $\rho_0$, decomposed into eigenstates we again denote
$\rho^{(t)} = \bar {\cal E}_{SV}^{(t)} (\rho_0)$. 
After iteration of the channel only eigenstates in the form of $\hat e_{0,kk}$ would remain 
in the density operator giving  
\begin{align}
\lim_{t \to \infty} \tr_A (\rho^{(t)})  = \sum_k \rho_{0,k} \ket{k}_F \bra{k_F} 
\end{align}
where the positive and real coefficients 
$\rho_{0,k} = \tr (\rho_0 {\hat I_d} \otimes \ket{k}\bra{k}_F)$ depend only 
on the initial Fourier amplitudes.    
The averaged channel is described as {\it dephasing}, as information about
superposition is lost.  However, the channel is dephasing in the Fourier basis, not 
in the vertex basis. 
 
We compute the probability of measuring a state to be at a particular vertex 
\begin{align}
p_v( \hat e_{0,kk}) &= \tr ( \hat P_v \hat e_{0,kk})\nonumber \\
& = \Big|\bra{v} \ket{k}_F \Big|^2  = \frac{1}{N} 
\end{align}
with $\hat P_v$ defined in equation \ref{eqn:Pv}. 
If there are no eigenstates with eigenvalue -1, 
upon iteration of the channel, the probability of vertex position measurements 
approaches a uniform distribution and in this sense the averaged quantum walk 
resembles a classical random walk. 
This is expected because the vertex and Fourier bases are mutually unbiased; 
$|\bra{v}\ket{k}_F | $ is independent of $v$ and $k$, for all $v, k$. 
After iteration of the channel 
vertex position measurements would not reveal any information about
the initial conditions.  However, the initial Fourier amplitudes 
are retained in the density operator and can be measured with other types 
of measurements such as measurements in the Fourier basis 
or those that measure correlations between vertex positions, as illustrated 
in more detail in section \ref{sec:acirc}.

In summary,  the cycle graph and hypercube are examples of a Cayley graph 
generated from an Abelian group.  For other finite Abelian groups, we can similarly 
generate a quantum random walk and a random walk with a Haar-random coin
using its Cayley graph.  
The averaged channel for the Haar-random coined quantum walk on the Cayley graph 
of Abelian group (corresponding to a depolarized coin) 
could have off-diagonal eigenstates (in the Fourier basis of the vertex subspace) 
with an eigenvalue of $-1$ and  that persist upon iteration of the channel. 
However, these non-decaying off-diagonal eigenvectors of the averaged channel are only present 
if all group elements (other than the identity) have even order.  
The indices of the Fourier components for the non-decaying off diagonal eigenvectors (in the form 
of equation \ref{eqn:ekkk}) of Fourier 
indices $k,k'$ must satisfy equation \ref{eqn:condj} for all digits of $k,k'$. 

If the averaged channel lacks eigenvalues that are $-1$, then the averaged 
channel is dephasing in the Fourier basis (of the vertex subspace) 
as all off-diagonal components in this basis decay under 
iteration of the channel.   In this case 
iteration of the averaged channel (of the quantum walk with a Haar-random coin operator) gives a uniform distribution in the probabilities of vertex measurements.  
The fully depolarized state $\tfrac{1}{dN} \hat I_d \otimes \hat  I_N$ is a fixed point of the channel. 
However because the channel is not ergodic, the iteration of the channel does not necessarily cause
the resulting density operator to approach this depolarized state. 
The amount of information about the initial conditions that is retained upon iteration 
of the channel is related to the number of fixed points of the channel. 
A measurement can reveal information about the relative amplitudes of the original 
state in the Fourier basis but not their original relative phases. 

\subsection{Measurements}
 
\section{Variances of vertex position measurements}
\label{sec:var}

In this section we estimate the size of fluctuations for measurements in the vertex 
subspace in a quantum walk with a Haar-random coin. 
Given a Hermitian operator $\hat B$ associated with a measurement, and 
a quantum state $\rho$, the conventional expectation value 
\begin{align}
\langle \hat B  \rangle = \tr (\rho\hat  B).
\end{align}

The density operator after applications of the Markov coin quantum channel $t$ times, starting 
from the state $\rho_0$ is 
\begin{align}
\rho^{(t)}_{\tilde V} = {\cal E}_{S\tilde V}^{(t)} (\rho_0). 
\end{align}
Here $\rho^{(t)}_{\tilde V}$ is a specific instance of the density operator 
resulting from $t$ different choices for the Haar-random coins
used during each step of the walk. 
 The density operator $\rho^{(t)}_{\tilde V}$ 
depends on $\hat V_1, \hat V_2, \ldots, \hat V_t$, 
the specific choices of the coin operators.  We subscript with tilde $V$ to make it clear 
that it depends on these choices.  
If we  carry out a particular quantum walk (specifying $\hat V_1, \hat V_2, \ldots, \hat V_t$) 
to give a density operator $\rho^{(t)}_{\tilde V}$,  
we would require multiple experimental copies of this same operator $\rho^{(t)}$, to 
 measure the expectation value $\langle \hat B \rangle  = \tr (\rho^{(t)}\hat B)$ in the lab. 
If we generate an ensemble of $\rho^{(t)}_{\tilde V}$ states using independently chosen 
coin operators, then we can compute the expectation value 
\begin{align}
\overline{ \langle \hat B  \rangle^{(t)} }  =  {\mathbb E}_{V_1, V_2 \ldots, V_t \in \mu_H}\left[ \tr (\rho^{(t)}_{\tilde V}\hat B)\right]. \label{eqn:eval}
 \end{align} 
This is the expectation value for measurements of $\hat B$ from an ensemble of random 
walks after $t$ iterations. 

Following equation \ref{eqn:EVVV},  
the expectation value of equation \ref{eqn:eval}, taking into account the probability distribution of the 
different randomly selected coin operators and after $t$ iterations of the quantum walk, is 
\begin{align}
\overline{ \langle \hat B  \rangle^{(t)} }  
& = \tr (\bar {\cal E}_{SV}^{(t)} (\rho_0)  \hat B) . \label{eqn:lk}
\end{align}
This depends on the averaged channel which was discussed in section \ref{sec:ave}. 
 
How do we describe the scatter or variance of the $\hat B$ measurements for an ensemble of walks?
Conventionally given quantum state $\rho$ and Hermitian operator $\hat B$ 
the variance of measurements 
\begin{align}
\sigma^2_{\hat B} &=  \langle \hat B^2 \rangle  - (\langle \hat B\rangle )^2 \nonumber \\
& = \tr ( \rho  \hat B^2 ) - \left( \tr (\rho \hat B) \right)^2.  \label{eqn:sigB}
\end{align}
When averaging over a distribution of different walks, we need to keep in mind that  
\begin{align}  
\left( {\mathbb E}_{\hat V_1 \hat V_2  \ldots \hat V_t \in \mu_H } \!\big[  \tr( \hat B \rho^{(t)} ) \big] \right)^2
  \! \! \ne 
{\mathbb E}_{\hat V_1 V_2 \ldots \hat V_t \in \mu_H} \! \Big[ \! \left( \tr( \hat B \rho^{(t)} )\right)^{\! 2}\! \Big].
\label{eqn:ne}
\end{align}
While the left hand side of equation \ref{eqn:ne} is straightforward to compute using the averaged channel with 
equation \ref{eqn:lk}, the right hand side is difficult to compute as it involves a second moment of the distribution of random unitaries. 
To get around this problem we estimate the change in expectation values caused by 
one iteration of the quantum walk 
\begin{align}
\delta \langle \hat B\rangle &=  \tr (  ({\cal E}_{S\tilde V}(\rho)   - \rho) \hat B) \nonumber \\
\delta \langle \hat B^2\rangle& =  \tr ( ( {\cal E}_{S\tilde V}(\rho)  - \rho) \hat B^2).\label{eqn:delta}
\end{align} 
With these changes we estimate a change in variance by linearizing equation \ref{eqn:sigB}
\begin{align}
\delta \sigma^2_{\hat B} &\approx \delta \langle \hat B^2 \rangle  - 2 \langle \hat B\rangle  \delta \langle\hat  B\rangle  \label{eqn:ds}
\end{align}
for a difference caused by one application of the channel. 
Inserting equation \ref{eqn:delta} into equation \ref{eqn:ds}, 
we estimate 
the change  in the scatter of $\hat B$ measurements 
\begin{align}
\delta \sigma^2_{\hat B} &\approx \tr (  ({\cal E}_{S\tilde V}(\rho) - \rho) \hat B^2)  \nonumber \\
& \ \ \ -  2\tr (   \rho B)
 \tr (  ({\cal E}_{S\tilde V}(\rho)  - \rho )\hat B). \label{eqn:ds2}
\end{align}
We compute the expectation value of the change in the variance 
\begin{align}
\overline{\delta \sigma^2_{\hat B}} & = 
{\mathbb E}_{\hat V \in \mu_h}[ \delta \sigma^2_{\hat B}]  \nonumber \\
&\approx 
   \tr ((\bar {\cal E}_{S V}(\rho)  - \rho) \hat B^2) 
  \nonumber \\
     & \ \ \   -   
    2 \tr (\rho \hat B) \tr ((\bar {\cal E}_{S V}(\rho)  - \rho) \hat B  ) . \label{eqn:Edisp}
\end{align}
This depends only on the averaged channel and is more easily computed 
than the expression on the right hand side of equation \ref{eqn:ne}. 

\subsection{The variance of vertex position estimates for the quantum walk on the cycle graph}

A property of the quantum walk with Hadamard coin on the line is that the variance in position measurements 
increases more rapidly than for the classical random walk \citep{Aharonov_1993}. 
In this subsection, 
we apply equation \ref{eqn:Edisp} to estimate the variance of position measurements in 
the quantum walk on the cycle graph with a Haar-random coin.

For a coined quantum walk on a line, an operator 
$\hat I_2 \otimes  \sum_{v =- \infty}^{\infty} v \ket{v}\bra{v}$ is used to measure the expectation value 
of position and its variance. 
On the cycle graph with $N$ vertices, 
we consider a similar operator that is Hermitian and diagonal in the vertex basis,  and periodic 
so that it has no large gaps in its spectrum; 
\begin{align}
\hat B = \hat I_2 \otimes \sum_{v=0}^{N-1} \tfrac{N}{2 \pi} \sin \tfrac{2 \pi v}{N} \ket{v}\bra{v}. 
\label{eqn:Bv}
\end{align}
For localized states,  the expectation value of this operator gives 
an estimate for the mean position index.  We use
a sine to be consistent with the cyclic nature of the graph, but keep in mind that  
states are labelled modulo $N$, so we allow vertex label $v$ to be negative.  
For small $v$, the value $\tfrac{N}{2 \pi}  \sin \tfrac{2 \pi v}{N} \sim v$. 
If $N$ is large and the system is begun in an initial state that is localized near $v=0$ then we can 
use $\hat B$ and the small angle approximation to measure the width of the vertex position 
 probability distribution. 

Expectation values for state $\rho$ and operator $\hat B$ given in equation \ref{eqn:Bv} 
\begin{align}
\tr (\rho \hat B )
& =\sum_{v=0}^{N-1} \rho_{vv} \tfrac{N}{2 \pi} \sin \tfrac{2 \pi v}{N}\nonumber \\
\tr (\rho \hat B^2 )  & =
\sum_{v=0}^{N-1}\rho_{vv} \left( \tfrac{N}{2 \pi} \right)^2 \sin^2 \tfrac{2 \pi v}{N}  \label{eqn:rr}
\end{align}
with notation 
\begin{align}
 \rho_{vv} &\equiv \bra{v} \tr_A( \rho) \ket{v}.
\end{align}
With the averaged channel for the quantum walk on the cycle graph with Haar-random coin in the form of \ref{eqn:ESV_cycle} we compute the expectation values
\begin{align}
\tr \Big(\bar {\cal E}_{SV,\text{cycle}}(\rho)& \hat B \Big) =\tr \Big[  \tfrac{1}{2} \! \sum_{vv'}  \Big(\! \ket{0}\bra{0} \otimes \ket{v+1}\bra{v'+1} 
\nonumber \\ 
& \ + \ket{1}\bra{1} \otimes \ket{v-1}\bra{v'-1} \Big) \rho_{vv'} \nonumber \\
&\  \times\hat I_2 \otimes  \sum_{v''}  \tfrac{N}{2 \pi} \sin \tfrac{2 \pi v''}{N} \ket{v''}\bra{v''} \Big] \nonumber \\
& =   \tfrac{N}{4 \pi} \! \sum_{v} \left( \sin \tfrac{2 \pi (v+1)}{N} + \sin \tfrac{2 \pi (v-1)}{N} \right)   \rho_{vv} 
\nonumber \\
\tr (\bar {\cal E}_{SV,\text{cycle}}(\rho) \hat B^2 )&=  \tfrac{1}{2}\! \left( \tfrac{N}{2 \pi}\right)^2 \sum_{v} \Big( \sin^2 \tfrac{2 \pi (v+1)}{N} \nonumber \\
& \ \ \ + \sin^2 \tfrac{2 \pi (v-1)}{N}\Big)   \rho_{vv} . \label{eqn:err}
\end{align}
Using equations \ref{eqn:rr} and \ref{eqn:err} 
\begin{align}
 \tr \Big( ( \bar {\cal E}_{SV,\text{cycle}}(\rho) &- \rho) \hat B\Big)=
 \tfrac{N}{4 \pi}\sum_{v} \Big( \sin \tfrac{2 \pi (v+1)}{N}  \nonumber \\
 &  + \sin \tfrac{2 \pi (v-1)}{N} - 2  \sin \tfrac{2 \pi v}{N} \Big)   \rho_{vv}
  \nonumber \\
\tr \Big( ( \bar {\cal E}_{SV,\text{cycle}}(\rho) &- \rho) \hat B^2 \Big)= 
 \tfrac{1}{2} \!\left( \tfrac{N}{2 \pi}\right)^2 \sum_v \Big( \sin^2 \tfrac{2 \pi (v+1)}{N}  \nonumber \\
& + \sin^2 \tfrac{2 \pi (v-1)}{N} - 2 \sin^2 \tfrac{2 \pi v}{N} \Big)   \rho_{vv}. 
\end{align}
If state $\rho$ mostly has power at small $v$ then  $\sin \tfrac{2\pi v}{N} \sim \tfrac{2 \pi v}{N}$ and 
\begin{align}
 \tr ( ( \bar {\cal E}_{SV,\text{cycle}}(\rho) - \rho) \hat B)& \sim 0 \nonumber \\ 
 \tr ( ( \bar {\cal E}_{SV,\text{cycle}}(\rho) - \rho) \hat B^2 )& \sim 1.
 \end{align}
 We insert these relations into equation \ref{eqn:Edisp} to estimate the increase in variance 
 \begin{align}
{\mathbb E}_{\hat V \in \mu_h}[ \delta \sigma^2_{\hat B}] &\approx 1. 
 \end{align}
 For an initially localized state, each iteration of the channel would on average increase 
 the variance of vertex position measurements by 1. 
 After $t$ iterations we expect the variance of vertex position measurements to be 
 \begin{align}
 \overline{\sigma^{2}_{\hat B}(t)}  \sim \sigma^2_{\hat B}(0) + t. \label{eqn:osigt}
 \end{align} 

For the classical random walk on the line and walkers initially begun at a particular location, 
the distribution of positions at time $t$ is Gaussian with a variance that 
is linearly dependent on $t$.   We recover the linear dependence of variance on iteration number for 
the quantum walk with the Haar-random coin on the cycle graph, 
but describing positive measurements for an ensemble of quantum walks. 
The linear dependence of variance on iteration number is the same as expected for 
the classical random walk.   The linear dependence of the variance 
is consistent with the recovery of classical 
behavior for a quantum walk on a cycle graph with a decoherent coin \citep{Brun_2003}. 

Even though information about initial conditions 
can be learned through measurements of correlations, decoherence caused by 
the Haar random coin causes the quantum walk to spread in the same way as 
the classical one.  The vertex position probability distribution for the Hadamard 
quantum walk and one instance of the Haar-random coined quantum walk
on the cycle graph is illustrated in Figure \ref{fig:walks} (see section \ref{sec:illust}).
Figure \ref{fig:walks}b shows that the variance in vertex position is approximately Gaussian for the quantum 
walk with the Haar-random, as would be expected from equation \ref{eqn:osigt}. 


\section{Summary and Conclusion}

In this manuscript we have explored 
a generalization of a quantum random walk on a $d$-regular connected simple graph, replacing the coin 
operator with a random matrix $\in U(d)$ that is independently drawn from a uniform 
distribution with respect to the Haar measure.  The random matrix is chosen 
independently each step of the quantum walk. 
Moments of Haar-random matrices (e.g., \cite{Mele_2024}) aid   
 in computing statistical properties of the Markov coin random quantum walk model.  
We find that the averaged channel, computed with the expectation value over randomly chosen 
operators operating on the coin subsystem,  depolarizes the coin subspace.  
Nevertheless, the channel itself is a unitary channel and 
 an initially pure state remains a pure state after each iteration of the walk.  
Expectation values of operators for ensembles of walkers can be computed with the averaged 
 quantum channel and its iterates.   

Even though the coined space is on average depolarized, we show that in some special 
cases, measurements of vertex positions can detect information about the initial conditions of the quantum 
state, even after repeated iteration of the averaged channel.   
The cases where this is  possible depend upon the graph.   
For a Cayley graph of an Abelian group, 
the initial state must contain off-diagonal terms in the Fourier basis,  
associated with eigenstates of the channel's superoperator with eigenvalue $-1$,  and these 
only exist if the order of all group elements (other than the identity) is even. 
Examples of such states are superpositions of two states of the Fourier basis. 
The existence of such eigenstates is related to 
conditions placed on a coined quantum random walk 
on the Cayley graph of an Abelian group for the vertex position distribution 
to approach a uniform distribution upon iteration of the walk \cite{Aharonov_2001}. 

We find that the averaged quantum channel of the Haar-random coined quantum walk 
has multiple fixed points and so is not ergodic.  For quantum walks 
constructed on a Cayley graph of an Abelian group, the number of linearly independent fixed points 
of the averaged channel is equal to the number of elements in the group. 
Because the fixed points do not decay upon iteration of the channel, information 
about an initial state is potentially measurable even after multiple iterations of the averaged channel. 
If there is no off-diagonal eigenstate with eigenvalue $-1$, then the averaged channel 
for the walk on a Cayley graph is {\it dephasing} in 
the Fourier basis, and the vertex position measurements approach a uniform distribution.  
However measurements of correlations between pairs of vertices can 
reveal information about the initial state, even after repeated iteration of the depolarizing 
averaged channel.  For example, 
the relative amplitudes of a state that is initially a superposition of two Fourier basis 
states, can be detected via measurement of correlations between two vertex positions in 
the averaged quantum channel.  

Even though the Haar-random coined quantum walk is a unitary quantum channel, 
an initially localized particle does not spread ballistically 
as is true in the Hadamard quantum random walk based on the cycle graph.  
We show this by estimating the scatter in position measurements for the Haar-random quantum 
walk on the cycle graph. 
This result is consistent with prior studies of an open quantum walk with a decoherent coin operator \citep{Brun_2003}.  
Decoherence is introduced by 
the independently chosen random matrices that are applied consecutively at each iteration of the quantum walk. 

Our Markov coin, because it is uniformly distributed according to the Haar measure,  
is straightforward and 
gives a nice example of correspondence between a quantum and classical walk and decoherence. 
Except for the special cases discussed above, 
the vertex distribution approaches the uniform distribution, as would 
a classical random walk with a fair coin.   Behavior associated with 
the classical random walk on a graph is embedded within the quantum walk via averaging over 
the Haar-random matrix distribution. 
In the case of quantum walks on Cayley graphs, the classical behavior is associated with dephasing with respect to the Fourier basis. 

Due to its decoherence,  
a quantum random walk with a Haar-random coin 
would probably not be useful in quantum search algorithms which are aided by ballistic 
spreading of the quantum state vector \citep{Childs_2004,Portugal_2018}.  
However, they may be useful in other contexts such as 
in measuring transport properties (e.g., \cite{Prosen_2015}) or modeling quantum systems 
with noisy subsystems or subsystems that interact with a thermal reservoir (e.g., \cite{Chisholm_2021,Ciccarello_2022,Stanzione_2025}).    

For future work, 
the case of a Haar-random coin could be generalized to coin operators drawn 
from other distributions.  More complex 
Markov models that perturb the system in position or coin or both spaces  (e.g., \cite{Bougron_2022,Derevyanko_2018}) could be studied.  
It might be possible to develop a theory for fluctuations exhibited by this simple model, 
improving upon our rough estimates in section \ref{sec:var}. 

\acknowledgements
ACQ is grateful for visitor support at the Max-Planck Institute for the Physics of Complex Systems (Dresden, Germany) May and June 2026. 
We enjoyed and benefited from helpful, illuminating and informative discussions with 
Akram Touil, Damian Sowinski, 
Abobakar Miakhel, Gabriel Alves, Felix Fritzsch, Gabriel Landi, Nathan Skerrett and Hans Singh. 

\bibliographystyle{elsarticle-harv}
\bibliography{walks}

\end{document}